\documentclass[sigconf]{acmart}
\AtBeginDocument{%
  }

\copyrightyear{2026}
\acmYear{2026}
\setcopyright{cc}
\setcctype{by}
\acmConference[ASSETS '26]{The 28th International ACM SIGACCESS Conference on Computers and Accessibility}{October 25--28, 2026}{Vila Nova de Gaia, Portugal}
\acmBooktitle{The 28th International ACM SIGACCESS Conference on Computers and Accessibility (ASSETS '26), October 25--28, 2026, Vila Nova de Gaia, Portugal}
\acmDOI{10.1145/3797867.3828997}
\acmISBN{979-8-4007-2521-0/2026/10}

\usepackage{booktabs}
\usepackage{tabularx}
\usepackage{multirow}

\begin{document}

\title{Supporting Allyship in Virtual Collaboration with Artificial Intelligence: A Scenario-Based Study}

\author{Crescentia Jung}
\orcid{0000-0003-4434-1723}
\email{cj382@cornell.edu}
\affiliation{%
  \institution{Cornell University}
  \city{New York}
  \state{NY}
  \country{USA}
}
\author{Ricardo E. Gonzalez Penuela}
\orcid{0000-0002-1344-3850}
\email{reg258@cornell.edu}
\affiliation{%
  \institution{Cornell University}
  \city{New York}
  \state{NY}
  \country{USA}
}
\author{Prashita Biswas}
\orcid{0009-0000-6009-9661}
\email{pb598@cornell.edu}
\affiliation{%
  \institution{Cornell University}
  \city{New York}
  \state{NY}
  \country{USA}
}
\author{Seungyoon Kwon}
\orcid{0009-0000-3434-1100}
\email{sk3448@cornell.edu}
\affiliation{%
  \institution{Cornell University}
  \city{New York}
  \state{NY}
  \country{USA}
}
\author{Shiri Azenkot}
\orcid{0000-0002-6701-4066}
\email{sa933@cornell.edu}
\affiliation{%
  \institution{Cornell Tech}
  \city{New York}
  \state{NY}
  \country{USA}
}
\renewcommand{\shortauthors}{Jung et al.}

\begin{abstract}
Recent research has revealed that accessibility in virtual collaboration is not only a technical problem but also depends on allyship: the informal, interpersonal practices through which people support others’ accessibility needs. Yet little research has examined how artificial intelligence (AI) might support allyship in collaborative accessibility contexts. To address this gap, we conducted a study where 18 participants (10 disabled, 8 non-disabled) reflected on four allyship scenarios with novel AI tools. Participants valued AI designs that they anticipated could help teammates express, interpret, and coordinate around access needs by reducing repeated disclosure and making allyship more actionable. At the same time, participants raised concerns when AI acted without consent, misrepresented users’ intentions, or displaced the interpersonal work of allyship. Based on our findings, we introduce allyship-support tools as a class of accessibility technologies, offer design guidelines, and discuss AI in particular as a means of supporting allyship.
\end{abstract}




\maketitle
\section{Introduction}
Virtual collaboration has become a fundamental part of how people work together. Tools like virtual meeting platforms, messaging tools, and shared documents allow team members to connect across distances and offer flexibility. However, virtual collaboration also introduces challenges, especially for people with disabilities. Prior work has documented how virtual collaboration creates barriers to participation, coordination, and access to shared information across a range of disability contexts. For example, studies with blind and low vision professionals highlight how screen sharing and visually oriented interfaces limit access to meeting content \cite{cha_you_2024}. Research with neurodivergent professionals points to challenges such as sensory overload and the need for flexible communication norms in remote meetings \cite{zolyomi_managing_2019, das_towards_2021}. These barriers can affect job performance, career growth, and sense of belonging on a team \cite{akter_beyond_2025, das_that_2024}.

In response to these barriers, researchers have long explored technologies to support accessibility. These technologies focus on improving individuals’ technical access to virtual collaboration, such as screen readers and captioning systems \cite{das_towards_2021, mcdonnell_social_2021, tang2021understanding}. Companies have extended these approaches through AI-powered features that automatically generate captions or describe visual content \cite{otter_ai_2026, zoom_ai_captions_2025, google_workspace_ai_meetings_2026, grammarly_2026}. While essential, these approaches focus on technical access to tools. However, research on mixed-ability teams suggests that accessibility is not only determined by technical access, but also shaped by interpersonal and relational dynamics \cite{zhao2026accessibility, mack_mixed_2021, jung_understanding_2026, yildiz_virtual_2023}.

In recent years, research has revealed that accessibility in teamwork is not just a technical problem, but depends on team norms, expectations, and practices, including how members share access needs, coordinate support, and take on accessibility-related labor \cite{mack_mixed_2021, alharbi_accessibility_2023, jung_understanding_2026}. In these settings, accessibility is not simply provided to individuals, but is collectively negotiated through ongoing interactions between disabled and non-disabled teammates \cite{bennett_care_2020, branham_collaborative_2015, wang_accessibility_2018, yildiz_virtual_2023}. Jung et al. \cite{jung_understanding_2026} further found that \textit{allyship}, the actions taken to support disabled team members, is foundational to enabling these practices. Allyship is enacted through everyday team interactions, such as recognizing access needs, offering support, and coordinating accommodations, that shape how accessibility is collaboratively negotiated over time \cite{jung_understanding_2026}. Despite this, little work has examined how technology might support the relational dynamics through which accessibility is enacted in collaborative work.

Meanwhile, artificial intelligence (AI) has shown promise to support collaborative work, such as for creative tasks and decision-making. Human-AI interaction research suggests that AI tools and agents can shape collaboration by helping teammates coordinate, communicate, and navigate group dynamics. For example, Suh et al. \cite{suh_ai_2021} found that AI tools shape social dynamics in collaborative settings by functioning as a co-creative mediator that seeds shared starting points, sustains creative momentum, and reduces interpersonal friction. Related work on autonomous AI agents further highlights how AI can influence group processes, as Johnson et al. \cite{johnson_exploring_2025} explored proactive AI agents as teammates that contribute to group discussions by introducing perspectives, facilitating coordination, and shaping team reasoning processes. This creates an opportunity to ask whether AI with agentic capabilities might also support allyship in teams, since it can identify opportunities for support in moment-to-moment interactions \cite{samrose_meetingcoach_2021, johnson_exploring_2025}. Yet, little work has examined how AI might support allyship in accessibility contexts.

To address this gap, we explore how AI might be designed to support allyship. We focus on AI because allyship often depends on recognizing when support may be needed, identifying what form of support may be appropriate, and coordinating responses as collaboration unfolds. AI with agentic capabilities may be able to support these moment-to-moment processes by surfacing relevant context, suggesting possible actions, or mediating coordination between teammates. Prior work demonstrates that this is a particularly challenging problem. Jung et al. \cite{jung_understanding_2026} found that although allyship is an important component of strong teamwork in mixed-ability teams, it is often difficult to enact in practice. For disabled teammates, disclosing access needs can carry professional risks, such as stigma and being perceived as less capable. At the same time, non-disabled teammates may want to help but remain uncertain about when support is needed, what form it should take, and how to offer it without overstepping. Together, these tensions make allyship a complex interpersonal process that directly shapes teamwork. Therefore, we ask: \textbf{What designs of AI can support allyship in virtual collaboration?}

We conducted a scenario-based study with 18 participants who collaborate in mixed-ability teams, including ten who identified as disabled and eight who identified as non-disabled. A scenario-based study allowed us to explore future designs of AI and surface participants’ expectations, critiques, and concerns \cite{carroll_scenario-based_2026}. We developed four scenarios depicting situations in which teammates gave or received allyship using novel AI tools in virtual collaboration platforms. We conducted 60-minute sessions with each participant over a virtual meeting platform. In each session, we presented the four scenarios one at a time; for each, the researcher first walked through the scenario and then conducted a semi-structured interview to understand participants’ reactions, perceptions, and preferences for designs of AI to support allyship.

Our findings suggest that AI was perceived as most promising when it helped teammates express, interpret, and coordinate around access needs while preserving human judgment and relational control. Participants valued designs that reduced repeated disclosure, surfaced opportunities to provide support, and distributed allyship-related labor across the team. However, participants were less receptive to AI that acted without consent, risked misrepresenting users’ intentions, or weakened the authenticity and interpersonal meaning of support. These findings suggest that AI can best support allyship not by automating it, but by making support more actionable while preserving agency, authenticity, and the relational nature of teamwork.

Building on these findings, we argue that supporting accessibility in virtual teams requires rethinking the role of technology: from a tool that provides individual access to one that scaffolds the interpersonal and organizational processes through which access is negotiated, enacted, and sustained. We propose \textit{allyship-support tools} as a class of accessibility technologies that scaffold how teammates recognize, communicate, coordinate, and respond to access needs within ongoing collaborative relationships. 

We make two contributions. First, we present findings from a scenario-based study with disabled and non-disabled participants examining how participants perceived possible AI designs for supporting allyship in virtual collaboration. This empirical contribution shows how participants anticipated AI might help teammates express, interpret, and coordinate around access needs, while also surfacing tensions around agency, authenticity, and control. Second, building from these findings, we introduce the concept of \textit{allyship-support tools} as a broader class of accessibility technologies. This conceptual contribution shifts the focus of accessibility from individual access to the relational processes through which support is enacted in teamwork. We offer design guidelines that emphasize actionability, consent, and the preservation of human judgment, offering guidance for how AI-supported systems, and allyship-support tools more broadly, might support allyship without displacing the interpersonal work that sustains inclusive teamwork.

\section{Related Work}
\subsection{Accessible Virtual Collaboration}
Virtual collaboration presents substantial accessibility barriers for people with disabilities. Prior work has identified challenges such as inaccessible screen-sharing interfaces, absent or unreliable captions, and shared documents that are incompatible with assistive technologies \cite{mack_mixed_2021, cha_you_2024, siebold_disability_2025, lacy_seat_2022, akter_beyond_2025, das_that_2024, kushalnagar_teleconference_2020, zolyomi_managing_2019, das_towards_2021}. To navigate these barriers, people with disabilities use assistive technologies and often undertake additional preparatory work such as requesting materials in advance, adjusting platform settings, or developing workarounds for inaccessible interfaces and documents \cite{pandey_understanding_2021, zolyomi_managing_2019, cha_you_2024, das_towards_2021, branham_collaborative_2015}.

However, these workarounds and preparatory work are still often insufficient because many accessibility barriers stem from the social organization of collaboration rather than from the software itself. A screen reader cannot compensate when a teammate shares visual content without describing it, and captions do not resolve the difficulty of following overlapping or unstructured discussion \cite{mack_mixed_2021, cha_you_2024}. As a result, accessibility in virtual collaboration often depends on what teammates do for one another. Prior work shows that teammates may support access by sharing materials before meetings, describing visual content, checking communication preferences, or making space to discuss needs \cite{mack_mixed_2021, wang_accessibility_2018, li_re-envisioning_2024, zhao2026accessibility}. These practices are often essential, but ad hoc and uneven, depending on whether teammates know about others’ needs, understand what support would be useful, and are willing to act \cite{mack_mixed_2021, siebold_disability_2025, jung_understanding_2026, zhao2026accessibility}. This points to the need for a more structured understanding of how accessibility can be supported in virtual collaboration.

\subsection{Allyship in Disability Contexts}
Allyship refers to actions taken to support or advocate for members of marginalized groups \cite{brown_what_2013}. In disability contexts, allyship often takes the form of interpersonal support rather than formal accommodations, which are typically institutional and policy-driven \cite{mack_mixed_2021, yildiz_virtual_2023, ostrove_one_2006, ostrove_reflections_2009, jung_understanding_2026}. Prior work on collaborative accessibility and interdependence similarly shows that access is negotiated through interaction between disabled and non-disabled collaborators rather than simply provided by one person to another \cite{bennett_care_2020, wang_accessibility_2018, branham_collaborative_2015}. These practices include describing visual content, checking communication preferences, and adjusting collaboration behaviors to support access needs \cite{mack_mixed_2021, wang_accessibility_2018, bennett_care_2020, jung_understanding_2026}. Yildiz and Subasi \cite{yildiz_virtual_2023} further describe how members of mixed-ability teams engage in collective access and ongoing dialogue about one another’s needs.

Jung et al. \cite{jung_understanding_2026} brought this conversation more directly into mixed-ability virtual teamwork, showing that accessibility practices shaped not only access, but also coordination, participation, and collaboration. They identified two conditions for effective allyship: disabled teammates must be able to express their needs, and allies need clear, actionable ways to respond. At the same time, prior work highlights tensions that can make allyship difficult to enact. Disabled individuals may hesitate to disclose disabilities or request support because of stigma, power dynamics, or concerns about being perceived as burdensome \cite{lindsay_systematic_2018, von_schrader_perspectives_2014, cha_you_2024, pandey_understanding_2021}, while allies may be unsure when and how to help without overstepping \cite{jung_understanding_2026, mack_mixed_2021}. Thus, prior work has established the importance of allyship for strong collaboration and identified conditions for its success or failure. However, it has not examined in depth how AI might support allyship.

\subsection{Technology that Supports Interpersonal Dynamics in Collaboration}
While there has been little research investigating allyship in HCI and CSCW, researchers have explored broader ways that technologies can support interpersonal dynamics in collaboration. One line of work examines how systems support people in supportive roles. For example, Lee et al. \cite{lee_supporting_2025} studied people who remotely supported family members and friends with mental health conditions and identified opportunities for technology to help them maintain communication, coordinate care, and manage emotional strain. Similarly, Wang et al. \cite{wang_practice_2025} examined peer counselors and identified opportunities for systems to help supporters better understand others’ situations and coordinate appropriate responses. However, this literature assumes relatively stable and unidirectional support roles, focusing on how supporters carry out their role, rather than on how allyship is negotiated between collaborators in reciprocal, ongoing relationships.

Since allyship is often enacted through everyday interpersonal dynamics such as communication, participation, coordination, and trust, a second relevant line of work examines how collaborative technologies shape these aspects of teamwork \cite{samrose_meetingcoach_2021, samrose_coco_2018}. For example, researchers have explored communication features that support more inclusive interaction, such as backchannels for additional communication during meetings \cite{mu_whispering_2024}, visual communication layers for coordinating attention and participation \cite{kim_we_2023}, and tools that support disclosure and help teammates better understand communication needs \cite{li_re-envisioning_2024}.

Building on this broader research on technologies that shape interpersonal dynamics in collaboration, recent work examines how AI, including agentic capabilities, can influence social dynamics in group work. This line of work is especially relevant as it shows that AI can affect how collaborators align, relate to one another, and navigate trust. Suh et al. \cite{suh_ai_2021} found that generative AI could sometimes act as a social connector by helping collaborators align quickly. At the same time, other work shows that AI can also introduce interpersonal tensions rather than simply improving connections. For example, Johnson et al. \cite{johnson_exploring_2025} found that proactive AI agents in synchronous teamwork raised concerns about trust, inclusivity, and social integration. Hwang et al. \cite{hwang_in_2024} showed that AI agents speaking on a person’s behalf raised concerns about authenticity, trust, and autonomy in social interaction. Fu et al. \cite{fu2024text} found that while users valued AI-mediated communication tools for reducing anxiety, they also expressed concerns about inauthenticity and the erosion of personal voice, particularly in close relationships. Prior work has explored proactive and context-aware AI systems that guide collaboration, but these have not been examined in the context of accessibility and allyship in mixed-ability teams.

This literature demonstrates that technology can shape how collaborators communicate, coordinate, and relate to one another. However, this work has primarily focused either on one-directional support within structured roles or on collaboration dynamics more generally. By contrast, allyship in mixed-ability teams is peer-based, reciprocal, and embedded in ongoing teamwork. Supporting it requires attention not only to coordination, but also to how teammates recognize access needs, decide how to respond, and negotiate the interpersonal meaning of support. Our work addresses this gap by exploring the design of AI to support allyship.

\section{Method}
To answer our research question, we conducted a scenario-based study to examine disabled and non-disabled people’s preferences for AI to support allyship. Our method draws on scenario-based design, which uses narratives of future use as low-fidelity design artifacts to explore how technologies may be experienced before they are implemented. Unlike interview studies, which primarily elicit reflections on past experiences, scenario-based studies present participants with hypothetical situations, allowing researchers to surface expectations, critiques, concerns, and values for technologies that do not yet exist but may be feasible in the near future \cite{carroll_scenario-based_2026}.

To operationalize this approach, we developed a set of hypothetical collaboration scenarios that depict how AI might be used in mixed-ability teams. Each scenario was presented as a storyboard illustrating a concrete collaboration situation together with a form of AI. These storyboards enabled participants to react to and reflect on potential uses of AI in the context of teamwork \cite{chang2024dynamic, luria2020social, seo2021impact, zimmerman2017speed, dangol2025want}. We first conducted a formative study to inform the development of our scenarios.

\subsection{Formative Study: Informing the Creation of Scenarios}
\label{formative_study}

Our formative study helped us prepare the main scenario-based study in two ways. First, it refined the problem space by identifying realistic allyship-related situations in mixed-ability teamwork that the scenarios should depict. These situations involved recurring interpersonal or coordination challenges and could plausibly benefit from AI. Second, it narrowed the design space by surfacing participants’ preferences for how AI should be represented, presented, and behave within those situations. Since the main study aimed to elicit participant reactions to AI support for allyship, the scenarios for the main study needed enough structure for participants to reason about specific design possibilities without becoming an open-ended design study.

To identify the problem space, we asked participants to reflect on allyship-related challenges they had experienced in mixed-ability teamwork and to describe situations in which AI might be valuable. Prior work has shown that accessibility in mixed-ability teamwork involves ongoing negotiation of support, including challenges in articulating access needs, determining when and how to offer help, and balancing support with respect for autonomy \cite{jung_understanding_2026}. We built on this work by using these previously identified challenges as a starting point for the formative study, probing how they appeared in participants’ lived experiences and whether there were additional allyship-related situations in which AI might be useful.

To explore how AI might support these allyship-related situations, we designed formative storyboards that varied how AI was represented, how visible its actions were to the team, and how proactively it behaved. We looked to research on AI with agentic capabilities in social and collaborative contexts. We focused on AI with agentic capabilities because allyship unfolds through moment-to-moment interactions \cite{jung_understanding_2026}. Unlike static or user-initiated features, AI with agentic capabilities can identify opportunities for support in context as collaboration unfolds \cite{samrose_meetingcoach_2021, johnson_exploring_2025}.

Prior work highlights several key dimensions: embodiment, visibility, and initiative. Embodiment defines how AI is represented, which can influence users’ sense of social presence and expectations. For example, virtual representations can enhance telepresence, while highly anthropomorphic forms may raise expectations that, if unmet, reduce perceived effectiveness \cite{nowak_biocca}. Visibility refers to how visible the AI’s interactions are within team interactions. Prior work shows that different levels of visibility, from subtle to overt, can shape trust and coordination by improving users’ understanding of the system’s behavior and intentions, as well as accountability for actions taken within the collaboration. However, increased visibility can also lead to over-reliance, misinterpretation, and added cognitive or social burden \cite{chan2024visibility, ehsan2021expanding, sauppe2015social}. Initiative describes the extent to which AI acts proactively versus passively; mismatches between expected and actual levels of initiative can undermine user trust and acceptance, and appropriate levels depend on task context and team dynamics \cite{norman2009design, meurisch2020exploring, liao2023proactive}. We used these dimensions to structure and narrow the design space.
\subsubsection{Participants and Procedure}
We recruited six participants (four disabled and two non-disabled) with experience working in mixed-ability teams, whose ages ranged from 21 to 44 years old. Participants were recruited through mailing lists, social media, and snowball sampling. Participants were invited to self-identify any disabilities, and disclosure was entirely voluntary. 

Each 60-minute session had two chronological parts. First, we asked participants to reflect on allyship-related challenges they had experienced in past teamwork and to describe situations in which AI might have been valuable. Second, we presented storyboards depicting different forms of AI to support allyship in virtual collaboration. These storyboards helped participants compare concrete versions of three design dimensions that would have been difficult to evaluate abstractly: embodiment (text-based versus anthropomorphic representations), visibility (private versus public interactions), and initiative (proactive versus passive behavior). These formative storyboards were not the same as the four scenarios used in the main study; rather, they were storyboards used to compare possible AI design dimensions, which informed the final scenario designs described in Section~\ref{final_scenarios}.

We used semi-structured interviews to elicit participants’ reactions, as well as examples of situations in which such systems might be useful. We analyzed the sessions through open descriptive codes, focusing on participants’ descriptions of preferences, concerns, and prior experiences that informed their reactions to the storyboards. 
\subsubsection{Findings}
While prior work has examined mixed-ability teamwork more broadly and surfaced allyship-related challenges \cite{mack_mixed_2021, alharbi_accessibility_2023, jung_understanding_2026}, our formative study specifically asked about allyship challenges. We asked about how participants experienced allyship-related breakdowns and where they imagined AI support could be valuable. Participants described difficulty articulating their accessibility needs and identifying appropriate forms of support. They also noted that requesting support could feel uncomfortable or exposing, particularly in group settings, highlighting the potential value of anonymity or mediated communication. In addition, participants described uncertainty about when and how to offer help, even when motivated to be supportive, indicating opportunities for systems that surface appropriate moments for intervention. Finally, participants emphasized tensions in how support is communicated, including concerns about overstepping, paternalism, or misinterpreting others’ needs.

In addition, participants expressed preferences regarding how AI should be integrated into teamwork, which shaped the final scenarios. They favored text-based interactions over anthropomorphic representations, private communication channels over public interventions, and systems that behaved passively unless users had specified preferences in advance. We set aside design alternatives such as more anthropomorphic AI representations, public-facing interventions visible to the team, and proactive AI behavior because participants expressed concerns that such design choices could heighten social risk or reduce user control.

Together, these findings helped us establish (1) the problem space by identifying which allyship-related situations are most appropriate for AI, and (2) design preferences for how AI should be represented, presented, and behave when supporting allyship. We elaborate on these in the following section.

\subsection{Scenarios}
\label{final_scenarios}
\begin{figure*}[t]
    \centering
    \includegraphics[width=0.95\textwidth]{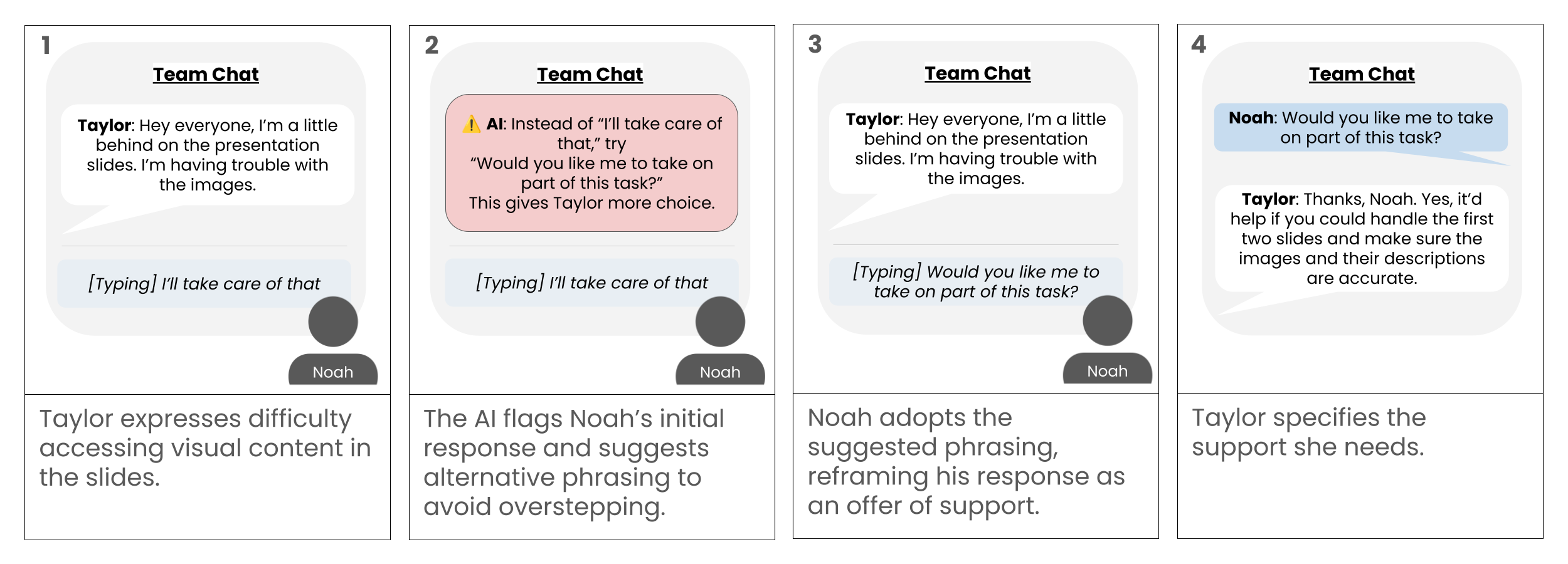}
    \caption{Example of Storyboard for Scenario 4: Phrasing Alternatives for Offering Help.}
    \Description{Four-panel storyboard with numbered panels, chat-style message bubbles, colored AI and user messages, and short captions beneath each panel. Taylor says she is behind on presentation slides and having trouble with the images. The AI flags Noah's draft response and suggests less presumptive phrasing. Noah revises his message, and Taylor specifies the support she wants. The panel captions describe this progression from disclosure, to AI intervention, to revised communication, to Taylor's clarification of needed support.}
 
    \label{fig:storyboard_example}
\end{figure*}
\begin{table*}[t]
\centering
\small
\setlength{\tabcolsep}{3pt}
\begin{tabular}{p{4cm} p{8cm}}
\toprule
\textbf{Scenario} & \textbf{Description} \\
\midrule

\textbf{1. Accessibility Profile Support} 
& Helps teammates articulate and share accessibility-related work preferences. The AI suggests possible accommodations and stores selected preferences in a profile teammates can reference. \\

\textbf{2. Anonymous Support Request} 
& Enables teammates to request support without direct exposure. The AI detects moments of need and offers options to route anonymous requests. \\

\textbf{3. Support Opportunity Prompts} 
& Identifies moments where support may be helpful and coordinates allyship. The AI suggests support options to teammates based on known preferences and helps organize follow-through. \\

\textbf{4. Phrasing Alternatives for Offering Help} 
& Supports reflective communication when offering help. The AI suggests alternative phrasings and explains their implications. \\

\bottomrule
\end{tabular}
\caption{Summary of AI assistance scenarios used in the study.}
\label{tab:scenarios}
\end{table*}

The scenarios were designed to reflect both the allyship situations and the AI design preferences identified in the formative study. Specifically, they address allyship challenges with preferred forms of AI, including text-based interaction, private visibility, and passive behavior unless preferences have been specified in advance. The AI tools depicted in the scenarios represented imagined future capabilities rather than existing systems. Figure~\ref{fig:storyboard_example} shows an example storyboard, and the full set of storyboards is included in Appendix~\ref{app:storyboards}. Table~\ref{tab:scenarios} provides a summary of the scenarios.

\textbf{Scenario 1: Accessibility Profile Support.} This scenario explored how AI might help make accessibility preferences clear. This design responds to our formative study finding that teammates sometimes struggled to articulate what kinds of support would be helpful and found examples useful for recognizing potentially helpful accommodations.

In the scenario, Alex, an autistic team member who experienced difficulty meeting deadlines, prepared to collaborate on a project. Rather than explaining his needs from scratch to each teammate, Alex interacted privately with an AI to create an accessibility profile. The AI used information Alex provided about his disability and work-related challenges to suggest possible support options, such as receiving reminders before deadlines, scheduling coworking sessions, allowing flexibility with deadlines, or enabling captions. Alex selected the options that reflected how he preferred to work. Only the preferences Alex selected were saved to his profile and made available to teammates, allowing them to understand his support preferences and consider how to respond when coordinating work together.

\textbf{Scenario 2: Anonymous Support Request.} This scenario explored how AI might mediate requests for support when directly asking a team member could feel uncomfortable or exposing. In the formative study, participants raised anonymity as a potential means of reducing discomfort and privacy concerns around requesting support, consistent with existing features in platforms like Zoom that allow users to request captions anonymously.

In the scenario, Casey was a new team member with ADHD. He attended a meeting where Jordan, the facilitator, shared multiple documents on the screen. Casey had previously indicated that receiving shared materials helped reduce cognitive overload. The AI used this preference and the meeting context to determine that the materials might be helpful. It privately asked Casey whether he wanted help requesting them. After Casey said yes, the AI offered several routing options, including asking the whole group, privately notifying a trusted ally, or asking the meeting facilitator. Casey chose to notify Jordan, and the AI prompted Jordan to share the links or email the materials without requiring Casey to ask directly in the group.

\textbf{Scenario 3: Support Opportunity Prompts.} This scenario explored how AI might help with establishing and coordinating teammates’ support preferences. Prior work has proposed technology-mediated reminders and coordination aids for accommodation-related support \cite{mack_mixed_2021}, while our formative study findings suggest that teammates may hesitate to intervene when unsure whether help is wanted or what form it should take, and that preferences might be used by AI to make suggestions. 

In the scenario, Riley was a team member with a chronic illness and ADHD. She had previously indicated in her accessibility profile that coworking sessions helped her with accountability. Riley shared in a team chat with Ethan and Sam that she was worried about forgetting to update a draft before the next day’s meeting. The AI recognized this as a possible support moment based on Riley’s message and saved preference. It first privately offered to set a reminder for Riley, which she accepted. The AI then privately prompted Ethan and Sam to consider offering a coworking session and disclosed that the suggestion was based on Riley’s stated preference. After Ethan and Sam agreed and offered the session to Riley, she accepted, and the AI scheduled it.

\textbf{Scenario 4: Phrasing Alternatives for Offering Help.} This scenario addressed the challenge where well-intentioned help can be perceived as overstepping or paternalistic \cite{jung_understanding_2026}. Our formative study participants described these challenges when offering support as an ally. Drawing from disability etiquette principles that emphasize autonomy and agency \cite{disabilityin_etiquette_guide, unitedspinal_top10_etiquette, cohen_disability_etiquette_epva, ndnrc_disability_etiquette, washington_accessibility_etiquette}, the design explores how AI might encourage more reflective communication among teammates.

In the scenario, Taylor was a blind team member who mentioned in the team chat that she was having difficulty with the images in a set of presentation slides. Wanting to help, Noah, a teammate, began typing, “I’ll take care of that.” Before he sent the message, the AI used Taylor’s message and Noah’s draft response to determine that Noah’s phrasing might unintentionally overstep by taking control of the task. The AI privately suggested alternative phrasings, such as, “Would you like me to take on part of this task?” or “Which images do you need help with?” It also explained that these responses might be more appropriate because they left room for Taylor to express what kind of support would be helpful. Noah then revised the message accordingly, which created space for Taylor to specify her preferences.

\subsection{Participant Recruitment and Demographics}
To be eligible for the study, participants had to (1) be part of a mixed-ability team that included both disabled and non-disabled members, (2) regularly use virtual meetings for teamwork, and (3) for disabled participants, self-report having experienced challenges in performing responsibilities due to their disability in virtual meeting contexts. These criteria ensured that participants had relevant, lived experience with accessibility and collaboration in virtual team settings, allowing us to ground our findings in real-world practices and challenges.

We recruited participants through mailing lists for people with disabilities and people interested in accessibility topics, social media sites, online groups for people interested in accessibility topics, posters on our university’s campus, and snowball sampling. Recruitment materials described the study as seeking people with experience working in mixed-ability teams that collaborate virtually. 

To determine eligibility, interested individuals first completed a survey in which we asked about their disability status, whether they worked on a team with both disabled and non-disabled members, and their use of virtual meetings. Participants were invited to self-identify any disabilities, and disclosure was entirely voluntary. Additionally, to respect privacy and considerations around disability status, we clarified that participants were not required to disclose the specific disabilities of their team members. Instead, participants were only asked to confirm whether their teams included members with disabilities. This approach allowed us to identify mixed-ability teams without requiring personal or detailed information about other team members’ disabilities. This study and all recruitment materials were approved by our university’s Institutional Review Board.

We recruited 18 participants (5 male, 13 female). Ten participants identified as having a disability (e.g., blind or low vision, attention-deficit/hyperactivity disorder (ADHD), anxiety, attention deficit disorder (ADD)). Eight participants identified as not having a disability. Participant ages ranged from 20 to 44 years (mean = 27.2, standard deviation = 5.7).

\begin{table*}[t]
\caption{Participant Demographics. We refer to non-disabled participants as N\# and disabled participants as D\#. }
\label{tab:participants}
\begin{tabular}{lllll}
\toprule
\textbf{ID} & \textbf{Age} & \textbf{Gender} & \textbf{Disability} & \textbf{Field} \\
\midrule
N1  & 34 & Male   & ---                             & Academia \\
D2  & 36 & Female & Blind                           & Marketing \\
D3  & 24 & Male   & Blind                           & Academia / Non-profit \\
D4  & 25 & Female & ADHD                            & Technology \\
N5  & 24 & Female & ---                             & Technology \\
D6  & 26 & Female & ADD                             & Academia / Technology \\
N7  & 26 & Female & ---                             & Academia \\
D8  & 24 & Female & ADHD                            & Education \\
N9  & 23 & Female & ---                             & Education \\
D10 & 29 & Female & ADHD                            & Creative \\
D11 & 20 & Female & Anxiety / slow processing speed & Education \\
D12 & 44 & Female & Blind                           & Technology \\
N13 & 24 & Female & ---                             & Education / Technology \\
N14 & 29 & Female & ---                             & Non-profit \\
D15 & 25 & Male   & ADHD                            & Academia \\
D16 & 29 & Female & ADD                             & Academia / Technology \\
N17 & 24 & Male   & ---                             & Technology \\
N18 & 24 & Male   & ---                             & Education \\
\bottomrule
\end{tabular}
\end{table*}

Table \ref{tab:participants} shows detailed participant demographics. We refer to non-disabled participants as N\# and disabled participants as D\#. 

\subsection{Procedure}
We conducted 60-minute scenario-based study sessions with each participant via a virtual meeting platform. Prior to the session, participants completed a demographic survey (e.g., age, gender, and current occupation). All interviews were audio-recorded, and participants received a \$30 gift card for their time. See Table~\ref{tab:participants} for participant details.

We used the same protocol for both disabled and non-disabled participants, since in mixed-ability teams, both disabled and non-disabled members may act as allies in different situations, and non-disabled participants may also experience temporary or situational needs. Using one protocol allowed us to examine how all participants understood, enacted, and responded to allyship across different roles, rather than treating disabled participants solely as recipients of support and non-disabled participants solely as allies. 

During the session, we presented four scenarios (see Section~\ref{final_scenarios}), one at a time. For each scenario, the researcher first walked the participant through the storyboard step by step and also described it verbally to ensure accessibility. This walkthrough introduced the collaboration context, the AI, and each panel of the storyboard. We then conducted a semi-structured interview about that scenario. For each scenario, the walkthrough and interview took around ten to fifteen minutes in total. We began with the same broad question, “What is your gut reaction to this design?” to capture participants’ initial reactions. We then asked questions from several categories depending on the scenario and the natural flow of the conversation. Below, we include examples of question categories.

\begin{itemize}
 \item \textbf{Perceptions and appropriateness of the design.} We asked participants how useful, appropriate, or supportive they found the design in the given situation, and in what kinds of team contexts it might or might not work well (e.g., ``Do you think this design would be helpful or appropriate here? Why or why not?'').

\item \textbf{Role-based perspectives and relational dynamics.} We probed how the design was understood from different positions in the interaction, including both offering and receiving support, and how interpersonal dynamics such as comfort, boundaries, stigma, or potential overstepping might shape these experiences (e.g., ``How would this feel for the person offering support? How about the person receiving it?'').

\item \textbf{Connections to past experience and team context.} We invited participants to reflect on similar situations from their own teamwork experiences and to consider how factors such as team norms, roles, responsibilities, and working relationships might influence how the design would function in practice (e.g., ``Have you been in a similar situation before? How might team norms or roles affect this?'').

\item \textbf{Coordination, accountability, and distribution of effort.} We explored how the design might shape teamwork processes, including how responsibility is assigned, how support is coordinated, and whether effort is fairly distributed across team members (e.g., ``How might this design affect coordination or responsibility within the team?'').

\item \textbf{Scenario-specific feature questions.} We also asked follow-up questions about design elements unique to each scenario. For example, if a scenario depicted a teammate receiving AI support, we asked how they would feel if they were in that teammate’s position or if that person were their teammate (e.g., ``If Riley were your teammate, how would AI nudges to others affect your confidence or hesitation to reach out?'').

\item \textbf{Disclosure and willingness to express access needs.} For scenarios involving the expression of access needs and preferences, we asked disabled participants additional questions about how the design might shape their willingness to disclose needs or request support. For example, in Scenario 1, which involved creating an accessibility profile, we asked, ``How might a design like this change how you think about expressing your needs?''

\item \textbf{Broader applicability of allyship and support.} We invited participants to reflect on allyship more broadly across scenarios, including situations in which non-disabled teammates might need support due to illness, caregiving responsibilities, or other temporary or situational challenges.
\end{itemize}

Throughout the session, we elicited details about participants’ prior experiences that informed their answers. For example, we asked questions such as, ``Can you recall a recent instance when…'' We probed for details about these experiences, including who was involved, what needs were addressed, what actions were taken, how others responded, and what challenges or tensions emerged.

\subsection{Data Analysis}
All sessions were audio-recorded and transcribed. During the sessions, the research team took notes and debriefed after each one to discuss initial observations. We analyzed transcripts from disabled and non-disabled participants together to examine allyship as a relational practice that emerges through interactions between teammates, rather than as separate experiences.

We conducted our analysis in two phases. In the first phase, two researchers performed descriptive open coding to summarize the data without interpretation \cite{saldana_2016}. During this process, we prioritized segments that included concrete examples of interactions or situations, excluding abstract or generalized statements about what participants believed they should do or felt socially expected to do. This helped us analyze how allyship was enacted in practice rather than described only in the abstract. The researchers also wrote analytical memos to note emerging observations. To ensure consistency and minimize bias, the two researchers independently coded the same transcript and reconciled discrepancies. They repeated this process with a second transcript to ensure alignment before dividing and coding the remaining transcripts independently. 

In the second phase, we conducted interpretive open coding to identify the concepts emerging from the data \cite{saldana_2016}. We followed the same process: two researchers independently coded the same transcript, met to discuss, repeated with a second transcript, and then divided the remaining transcripts. We adopted this two-phase approach to first identify and focus on concrete accounts of participants’ experiences, and then develop interpretations of participants’ experiences. 

After all transcripts were coded, we conducted a cross-case comparison analysis \cite{miles2014qualitative} to identify patterns across participants. We constructed a matrix with 26 interpretive codes as rows and 18 participants as columns. In each cell where a participant was associated with the code, we described how the code manifested for that participant. This allowed us to examine how each construct manifested across participants, identify variations and edge cases, and surface broader themes. 

We then reviewed the matrix to identify higher-level patterns and relationships across codes, grouping related constructs into broader thematic categories that informed our two high-level findings themes: (1) reducing the labor of expressing, interpreting, and coordinating access needs, and (2) preserving human judgment, control, and team dynamics. Throughout the analysis, the research team met regularly to discuss emerging themes and maintain alignment. Our analysis resulted in two higher-level themes, six second-level themes, and 26 interpretive codes. For example, we applied the code “AI mediation as threat to authenticity” to the quote, “If Riley later gets to know that [her teammate] was prompted by an AI to do that, she might feel that it’s coming from an inauthentic place” (N7). A summarized version of the matrix with themes is provided in Tables~\ref{app:cross_case_summary_part1} and~\ref{app:cross_case_summary_part2} in Appendix~\ref{app:cross_case_matrix}.

\section{Findings}

\subsection{Reducing the Labor of Expressing, Interpreting, and Coordinating Access Needs}
\subsubsection{Addressing Breakdowns in Interpreting Needs and Responses}
Both disabled and non-disabled participants, but especially non-disabled participants, wanted to support others but struggled to determine what someone needed or whether intervening would be appropriate at all. This uncertainty was often tied to concerns about overstepping or appearing patronizing. For example, N14 worked for several years at a global human rights organization on a large distributed team. She explained that \textit{``sometimes it doesn't feel like there's a forum to ask what people might need,''} making it difficult to clarify needs directly. In the absence of clear information, she relied on inference, which could lead to misdirected or unwelcome support. She described a situation where she inferred, based on a colleague’s behavior, that he might be neurodivergent, and began sending him deadline reminders as a form of support during project coordination. She later learned that he used a wheelchair due to a mobility disability, meaning her accommodation did not address his actual needs. As she put it: \textit{``I sort of started accommodating him by texting him reminders... and I was like, oh no, he's in a wheelchair. Not what I thought at all.''} Reflecting on Scenario 1 (Accessibility Profile Support), she noted that a structured way to communicate needs could help avoid this kind of misinterpretation by making appropriate forms of support more explicit.

When reacting to Scenario 3 (Support Opportunity Prompts), non-disabled participants noted that AI-generated suggestions would be difficult to act on without sufficient context about why support was needed or whether it would be welcome. Similarly, for Scenario 4 (Phrasing Alternatives for Offering Help), N18 was unsure how to offer help without overstepping, worrying about sounding overly benevolent. N17, a software engineer on a fast-paced development team, described this uncertainty through an interaction with his manager, who had a motor disability. When N17 asked him to quickly log in and fix an issue, his manager explained that he could not type as fast. N17 was unsure how to respond: \textit{``I did not know if I should apologize for what I had asked him to do, or if I should be sorry for his situation, or if I should not, because I thought that would be insensitive.''} He also hesitated to ask more questions, unsure whether that would feel caring or intrusive. 

At the same time, disabled participants described suppressing needs in order to avoid negative judgments from teammates. For Scenario 1 (Accessibility Profile Support) and Scenario 2 (Anonymous Support Request), participants noted that making needs publicly visible could threaten perceptions of competence or professionalism. D10, an art director in the advertising industry with ADHD, described difficulty keeping up with fast-paced meetings, including tracking multiple points of discussion, prioritizing information, and identifying next steps. To manage this, she relied on AI tools such as meeting transcription and summarization systems that generated notes and action items after meetings. However, she hesitated to ask teammates for support, explaining, \textit{``I know I have ADHD, but sometimes I don't want to let them know I have ADHD, even though I need a little bit of support and help.''} Instead of requesting clarification or additional support from teammates, she used AI privately to compensate for these challenges. This meant that potential forms of interpersonal support, such as teammates slowing down, sharing materials, clarifying priorities, or checking in on understanding, were not enacted. Her decision to rely on AI independently reduced the need for disclosure, but also limited opportunities for teammates to recognize and respond to her needs.

In addition to challenges in identifying and sharing needs, disabled participants sometimes failed to recognize whether they needed support even when they were already implementing accommodations and strategies. D16, a software engineer with ADD, already relied on tools such as Pomodoro timers to stay focused and task trackers and post-it notes to keep herself accountable. However, she still felt uncertain about what additional support might be helpful in fast-paced collaboration. She reflected, \textit{``subconsciously I think I do need help, I just don't know what help I need.''} Despite actively using multiple strategies to manage attention and memory-related challenges, she described lingering uncertainty about what further support might be useful, particularly in collaborative settings where she needed to keep up with responsibilities, follow through on tasks, and remain accountable to teammates.

\subsubsection{Reducing the Burden of Expressing and Coordinating Access Needs}
In response to Scenario 1 (Accessibility Profile Support), participants emphasized that the profile could reduce the repeated labor of explaining preferences, needs, or accommodations to different collaborators over time. This was not only raised by disabled participants, but also by non-disabled participants who compared the design to other situations in which their teammates had to repeatedly clarify how they worked, communicated, or wanted to be understood. 

For example, N13 compared the accessibility profile to existing profile features in Slack, such as name pronunciation, which help teammates know how to address someone correctly without requiring repeated explanation. She also drew a parallel to a manager she had worked with who routinely told colleagues at the start of meetings that her note-taking helped her stay focused and did not mean she was disengaged. Drawing on these examples, N13 suggested that an accessibility profile could similarly make interaction preferences or support needs visible up front, so teammates would better understand how to work with someone without needing repeated clarification.

Disabled participants also shared similar sentiments. D10 emphasized the value of not having to repeatedly explain her needs to others, noting, \textit{``Rather than explaining my situation to each person individually, I think this function is really helpful.''}

Beyond reducing repetition, both disabled and non-disabled participants also described how making needs persistently visible could enable more direct and effective support. N18 explained that clearly stated preferences made it easier to respond appropriately: \textit{``it makes it really explicit, allows me to give the kind of support that the person needs in a clear-cut manner, without the need to sidestep around the issue.''} In this sense, participants anticipated that profiles could not only store information, but also translate needs into actionable forms of support.

Preset support options also carried social meaning for disabled participants. D16 explained that seeing these options signaled that accessibility was already normalized within the team or organization, influencing whether she felt comfortable selecting them. 

\subsection{Preserving Human Judgment, Control, and Team Dynamics}
\subsubsection{Distributing the Labor of Allyship}
Non-disabled participants shared how they noticed that the work of tracking, coordinating, and responding to access needs was often concentrated in a small number of individuals, rather than distributed across the team. When reflecting on Scenario 1 (Accessibility Profile Support), N13 described her experience working on a fully remote software development team with disabled team members. For example, one teammate she worked with closely had ADHD. The teammate had accessibility needs, including shorter meetings, access to captions or transcripts, and receiving materials and summaries to support following discussions.

Responsibility for coordinating her team’s needs was largely taken on by the project manager, who served as the primary point of contact. The project manager helped adjust meeting structures, shared materials in advance, and ensured that notes and action items were distributed after meetings. While this support benefited the team, N13 noted that it became increasingly difficult for one person to manage as team demands grew. Following a company merger, meetings became longer and required coordination across multiple teams, increasing the complexity of maintaining these practices. As she explained, \textit{``[our project manager] was kind of the point of contact… I did notice that at some point she was having more trouble accommodating all needs… she’s the one trying to do all the things that this tool would handle, but herself.''} N13 found value in AI that can help track accessibility needs over time and support their coordination across the team, rather than relying on one person to manage them alone.

Non-disabled participants also emphasized that an ally’s ability to provide support was shaped by their capacity, resources, and role within the team. In response to Scenario 3 (Support Opportunity Prompts), participants pointed to the value of distributing responsibility by prompting multiple teammates when support might be needed, rather than relying on a single attentive individual. N1 described this limitation from the perspective of a team member without formal responsibility for coordination. He explained, \textit{``I might not have the support myself (as an ally) to be able to accommodate those people,''} particularly in situations where multiple support needs arise at once or when he lacks the time or resources to respond to them. He highlighted that even well-intentioned allies may lack the time or bandwidth to consistently take on this work.

Similarly, N7 described how the support she provided depended on her role within the same broader team context. In her lab, which included members and collaborators with a range of disabilities, including motor impairments, low vision, blindness, and hearing-related access needs, she participated both as a collaborator and as an organizer of a seminar series. In her organizer role, she viewed accessibility as part of her formal responsibility. For seminars conducted over Zoom, this included proactively ensuring that captions were enabled, sharing materials ahead of time, and checking with attendees about their usual accommodations and any additional support they might need. As she explained, she would \textit{``100\% reach out to everybody [attending the seminar].''}

In contrast, when participating as a fellow lab member or collaborator, N7 did not see herself as the primary person responsible for coordinating accessibility, especially when someone else, such as the meeting facilitator, was already responsible for those needs. In those situations, she described taking a less proactive role by default, explaining, \textit{``I don't personally feel liable to go an extra mile unless somebody's needs are not being met satisfactorily.''} N7’s responsibility for accessibility was shaped by team role: she would step in when unmet needs became apparent, but did not see herself as the person who should initiate accommodation efforts when that responsibility already belonged to someone else.
\subsubsection{Accounting for Organizational and Social Constraints}
Disabled participants emphasized that allyship depended on the broader organizational and social context in which collaboration occurred. These constraints shaped both those requesting support and allies in distinct ways, influencing whether and how allyship could be enacted.

When reflecting on Scenario 1 (Accessibility Profile Support) and Scenario 2 (Anonymous Support Request), disabled participants described how concerns about professional risk influenced their willingness to make needs visible, even when such tools reduced the effort of requesting support. D6 was a part-time technical program manager and a graduate student with ADD. She did not use formal accommodations at work, because asking for support could feel risky in workplace contexts, as it might shape evaluation.

\textit{\begin{quote}
``I've felt hesitant from my manager to ask for support because I was afraid it would come up in a performance review... they'll use any excuse... It's the fear of this being used by HR as a reason to retaliate against an employee.''
\end{quote}}

For D6, tools that made needs more visible were only useful if the surrounding environment felt safe. She contrasted her corporate work environment with academia, where she felt the same tools would be \textit{``super helpful and not used against me.''} These accounts suggest that AI designed to support disclosure or requests would depend on users trusting that it will not negatively affect how they are perceived or evaluated.

Participants also described that whether AI-surfaced opportunities for support resulted in follow-through depended on team size. In Scenario 3 (Support Opportunity Prompts), surfacing a need did not necessarily guarantee follow-through when responsibility was diffuse. D4, a graduate student with ADHD who had experience in both small project teams and larger class-wide collaborations, explained that in larger teams there was often a sense that \textit{``someone will get to it.''} In relation to Scenario 3, she felt that even when it was visible to the whole team that someone needed access to shared materials, no specific individual would feel responsible for responding. By contrast, D4 felt that in smaller teams, someone was more likely to recognize that they should step in and follow through, rather than assume that another person would handle it.

\subsubsection{Preserving Control, Authenticity, and Human Judgment} 
\label{preserving_control}
Non-disabled participants worried that when AI mediated too much of an allyship interaction, it could weaken the intention, personal expression, and interpersonal learning that made allyship meaningful in the first place. For many, support was not only about whether help occurred, but its provenance. The support would be welcomed if it were recognizably grounded in another person's own care, judgment, and effort.

This concern appeared in responses to Scenario 3 (Support Opportunity Prompts). N7 explained that if Riley later learned a teammate had only acted because of an AI prompt, the support might feel less genuine: \textit{``If Riley later gets to know that [her teammate] was prompted by an AI to do that, she might feel that it's coming from an inauthentic place.''} D11 raised a related concern as an ally, explaining: \textit{``I'd be like, ‘Oh, does this person not feel comfortable reaching out to me directly (for help)?’ And that might worry me a little bit.''} Participants valued the interpersonal act of offering and receiving support as much as the support itself, not just the outcome of the help. Mediation by AI raised questions about whether support reflected the ally's own intention or merely compliance with a prompt. D3 similarly emphasized that \textit{``The best role technology can play is making the steps of interaction a little bit easier, instead of taking that very human-to-human interaction out.''}

Both disabled and non-disabled participants expressed similar concerns about AI mediating how support was communicated. In reactions to Scenario 4, they worried that AI-generated rephrasing could misrepresent tone and intentions, reducing control over self-expression or introducing biased interpretations. N14 worried that repeated rephrasing could create \textit{``a warped idea of the kind of person you are,''} gradually shifting how someone came across. D16 likewise emphasized that \textit{``when asking for someone's help, the delivery matters. I would personally write messages in my own tone and in my own intentions.''} N17 said AI-suggested phrasing made exchanges feel \textit{``not natural anymore,''} while N18 cautioned that such tools would depend heavily on their sensitivity to social nuance and could otherwise come across as \textit{``really blunt.''} 

Participants also questioned whether AI-mediated support would help people grow into better allies over time or simply substitute for that process. D11 explained:

\textit{\begin{quote}
``If you're too reliant on AI being able to say things that you're not as comfortable saying to the group, then maybe you'll unintentionally train yourself to not ask questions outside of [using the AI].''
\end{quote}}

N17 similarly reflected on a difficult conversation with his manager, who had a motor disability. The interaction was difficult for N17 because he needed to respond appropriately after learning about his manager’s needs, without clear guidance on what would be respectful. He explained, \textit{``I was way more comfortable in the later conversation that we had, and I could feel that even my manager was way more open and comfortable having those conversations with me.''} For him, learning came from working through the initial discomfort and uncertainty, rather than avoiding the interaction altogether. N18 raised a broader concern, whether AI risked \textit{``only performatively helping allyship''} without fostering substantive change in how teammates relate to one another.

\subsubsection{Supporting Coordination Without Overstepping Social Boundaries}
Both disabled and non-disabled participants preferred AI for low-stakes coordination, but were less comfortable when it acted on matters with greater interpersonal or logistical consequences. This distinction appeared in responses to Scenario 3 (Support Opportunity Prompts). N7 explained that she would be comfortable with AI supporting simple accommodations, such as prompting someone to turn captions on, because these actions felt easy to carry out and did not require much personal commitment. However, she felt differently when the requested support imposed a more substantial demand on the other person’s time or schedule:

\textit{\begin{quote}
``If AI is reaching out to turn captions on, I'd be comfortable with that. But something which requires me to block time on my own calendar... for those kinds of events, I would feel more comfortable directly talking to Riley.''
\end{quote}}

N7 distinguished between forms of support based on how much time or effort they asked of others. She viewed low-effort actions, such as turning on captions, as appropriate for AI coordination because they required little time or negotiation. By contrast, she felt that requests that involved committing time carried greater interpersonal weight, making direct communication preferable so that availability and expectations could be negotiated together. D12 drew a similar distinction. She favored AI for low-stakes actions like sending a document. In contrast, more demanding or socially weighty requests should be handled through direct interpersonal communication.

Participants were cautious about AI taking action without explicit, situational consent. In Scenario 3, this concern centered on the boundary between AI suggesting support and coordinating it. For example, participants were generally more comfortable with AI offering Riley a reminder after she posted that she might forget to update the draft, because this kept the decision with her. They were more hesitant when AI moved beyond this and contacted teammates about arranging a coworking session or scheduling that session on the calendar, since these actions involved other people and could commit Riley to support she might not want in that moment. D12 explained, \textit{``I would want the control to be with the person that's needing the help… I wouldn't want AI to assume all the time that it should go ahead and schedule something.''} Participants wanted the disabled teammate to retain control over whether support was initiated in a given moment, rather than having AI act automatically based on a saved preference. D15 similarly emphasized that both the person requesting support and the ally should retain agency, including whether support is requested, offered, or accepted.

\section{Discussion}
We found that participants valued designs that they anticipated could reduce repeated disclosure, surface opportunities to provide support, and distribute allyship-related labor across the team. However, participants were less receptive to AI that risked misrepresenting users’ intentions or weakened the authenticity and interpersonal meaning of support. Together, these findings suggest that AI may best support allyship not by automating it, but by scaffolding recognition, coordination, and reflection while preserving agency, authenticity, and the relational nature of teamwork. Building on our findings, we argue that supporting accessibility in virtual collaboration requires a different technological framing: allyship-support tools. We outline design guidelines for these tools and examine the considerations for AI in this space.

\subsection{Toward Allyship-Support Tools as a Class of Accessibility Technologies}
\begin{table*}[t]
\centering
\caption{Comparison of assistive technologies, helper-mediated accessibility systems, and allyship-support tools.}
\label{tab:allyship-support-tools}
\small
\setlength{\tabcolsep}{3pt}
\begin{tabular}{p{2cm}p{3.2cm}p{3.2cm}p{3.2cm}}
\toprule
& \textbf{Assistive Technologies} & \textbf{Helper-Mediated Accessibility Systems} & \textbf{Allyship-Support Tools} \\
\midrule
\textbf{What It Does} 
& Supports an individual user’s access directly 
& Supports accessibility by involving another person in performing the needed work 
& Supports how teammates identify, communicate, coordinate, and respond to access needs \\

\textbf{Typical Logic} 
& The system performs or automates accessibility work for one user 
& Accessibility is achieved by routing work to a helper 
& The system scaffolds team-based support and coordination around accessibility \\

\textbf{Examples} 
& Screen readers, captions, AI-generated descriptions, interface adaptations 
& Crowdsourcing, friendsourcing, human-in-the-loop systems 
& Accessibility profiles, anonymous request features, coordination aids, reflective communication support \\

\textbf{Primary Unit of Support} 
& Individual 
& One or more distributed helpers 
& Team relationship / mixed-ability collaboration \\
\bottomrule
\end{tabular}
\end{table*}
Accessibility research has emphasized two models of support: assistive technologies that help an individual user interact with inaccessible systems \cite{wobbrock_ability-based_2011, cha_you_2024, zolyomi_managing_2019, das_towards_2021, li_re-envisioning_2024}, and human-mediated accessibility systems such as crowdsourcing and friendsourcing \cite{lasecki2013answering, brady2013investigating, brady2013visual, brady2014friendsourcing, burton2012crowdsourcing, guo2016vizlens, gleason2016vizmap, brady2015using, bigham2010vizwiz}. Although both have substantially advanced accessibility, neither fully captures how support operates in teamwork, where accessibility often depends on how teammates recognize needs, coordinate responses, and negotiate support within ongoing relationships.

Prior work on collaborative accessibility and interdependence has shown that access is often relational and co-constructed through interactions among people, technologies, and environments \cite{bennett2018interdependence, branham_collaborative_2015, wang_accessibility_2018, yildiz_virtual_2023}. We build on this foundation, but focus on a more specific collaborative problem. In our findings, the central challenge was not only that access was shared, but that collaborators had to determine when support was needed, how to offer it without overstepping, how to disclose needs without incurring stigma, and how to distribute accessibility-related labor across the team. In this context, allyship names the collaborative practice through which this relational accessibility is enacted.

We therefore propose \textit{allyship-support tools} as a class of accessibility technologies. Unlike traditional assistive technologies, these tools are not designed solely to help one disabled user access content or interfaces. Unlike crowdsourced or friendsourced systems, they do not primarily route accessibility work to another person to complete on the user’s behalf. They also differ from technologies designed for caregiving or counseling, which generally assume more stable and asymmetrical support roles \cite{lee_supporting_2025, wang_practice_2025}. Instead, allyship-support tools scaffold the relational and coordinative work around support while keeping the people involved central to the interaction (see Table~\ref{tab:allyship-support-tools}).

At the same time, framing these systems explicitly as accessibility tools may not always be desirable. Several participants described wanting support without necessarily disclosing a disability, suggesting that allyship-support tools should also accommodate people who prefer to express work preferences, communication needs, or support preferences without labeling them as disability-related. Existing non-AI practices already support some of this work, including “work with me” statements \cite{buch_workwithme_2020, maymin_howtoworkwithme_2020}, personal user manuals \cite{atlassian_user_manual_nd}, and team working agreements \cite{scrum_team_working_agreement_nd}. These tools allow people to describe how they collaborate best, what kinds of support are useful, and how teammates should communicate or coordinate with them without requiring disclosure. Connecting allyship-support tools to these broader practices may make them useful for a wider range of interpersonal needs, including temporary illness, caregiving responsibilities, or other situational constraints. However, broadening the framing in this way introduces a tension. When tools describe access needs in general terms, disability-specific barriers can go unnamed and unaddressed, and teams lose the moments of explicit discussion through which they learn about and take ownership of accessibility.

This distinction also points to when AI may be useful versus when it may overstep. Participants’ reactions suggest that AI may be more appropriate for low-stakes coordination that depends on timing or context, such as reminding teammates of an expressed preference, helping route a request for materials, or prompting an accessibility action that might otherwise be forgotten. By contrast, participants were more cautious when AI moved into socially weighty actions, such as committing someone’s time, initiating support without situational consent, or mediating interactions that required direct interpersonal negotiation. Thus, systems should not simply add AI to allyship-support tools wherever support is possible. Designers should consider whether a particular use case requires AI’s contextual and coordinative capabilities, or whether it can be better handled through non-AI interface and workflow design. Taken together, allyship-support tools should support flexible self-presentation and lightweight coordination while preserving opportunities for teams to recognize, discuss, and take responsibility for access.

Our findings also demonstrate why these tools should not simply be replaced by more capable automation (as discussed in the following section). Participants worried the AI in our study would replace human expression, weaken authenticity, or remove opportunities for interpersonal learning. This suggests that the limitation is not simply that current AI remains imperfect. Rather, some aspects of accessibility support are fundamentally relational: who offers help, how that help is offered, and whether it reflects care and judgment all shape how allyship is experienced. In this sense, allyship-support tools are not only tools for accommodation, but are technologies intended to help build trust, clarify needs, and distribute responsibility in teamwork.

These concerns connect our findings to longstanding critiques of technosolutionism, the assumption that complex social problems can be resolved through technical intervention \cite{lindtner2016reconstituting, mackay1999paper}. Allyship is not a task to be completed but a relational practice sustained through ongoing negotiation, and our participants resisted framings in which AI would simply automate it. As demonstrated in Section~\ref{preserving_control}, participants worried that AI mediation could weaken the authenticity of support and remove the very moments of discomfort and miscommunication through which teammates learn to support one another. N17's account of a difficult first conversation with his manager, an interaction he described as becoming more comfortable over time, illustrates how friction itself can be generative. Rather than treating such friction as inefficiency to be smoothed away, our findings suggest it can be a site of learning: moments of uncertainty or awkwardness help teammates negotiate expectations, surface unstated preferences, and build shared responsibility for access over time \cite{romerolauro_resilience_2026, kapur_productive_2008}. This reframes the design goal from eliminating friction to preserving productive friction, ensuring that allyship-support tools scaffold these interactions without displacing the interpersonal work through which allyship develops.

\subsection{Design Guidelines for Allyship-Support Tools}
\textbf{Design for actionability, not visibility alone.}
Prior work on accessibility in collaborative settings has emphasized the importance of making access needs visible and discussable, whether through disclosure, shared communication norms, or explicit discussion of support needs \cite{mack_mixed_2021, li_re-envisioning_2024, jung_understanding_2026, yildiz_virtual_2023}. However, our findings suggest that visibility alone is insufficient for enabling effective support. Even when teammates were aware that someone might need support, allyship often stalled because collaborators remained uncertain about what action would actually be helpful, appropriate, or welcome. In this sense, our findings extend prior work on collaborative accessibility by identifying a gap between awareness and response: knowing that support may be needed does not necessarily make that support actionable \cite{mack_mixed_2021, wang_accessibility_2018, jung_understanding_2026}. This reframes the design problem from one of disclosure to one of awareness and coordination. Allyship-support tools should therefore do more than surface needs; they should help convert those needs into concrete, context-sensitive forms of action. Examples include structured support preferences, examples of accommodations others have found useful, and prompts that clarify what kinds of responses would be helpful in a given situation. 

\textbf{Reduce coordination burden while preserving user control.}
Prior research on accessible collaboration and AI-assisted systems has shown that technology can reduce the burden of repetitive support work, including disclosure, coordination, and reminders \cite{lee_supporting_2025, wang_practice_2025, jung_understanding_2026}. Our findings revealed similar insights, but an important limitation: participants were hesitant when systems acted autonomously without explicit, situational consent. This extends prior work on AI proactivity by showing that, in accessibility contexts, initiative is not only a matter of efficiency or convenience, but also one of agency, disclosure, and social risk \cite{johnson_exploring_2025}. A system that oversteps may expose someone in ways that feel unsafe, misrepresent their preferences, or undermine their control over how they are perceived by others. These risks are shaped by workplace power dynamics: an accessibility profile or AI-generated prompt may be experienced very differently depending on who can see it, and a support preference shared among peers carries different stakes than the same preference visible to a manager or accessible to HR. D6's concern that disclosed needs could surface in a performance review illustrates how visibility that feels safe among peers can become a professional risk when it extends up a reporting hierarchy. Consent, therefore, cannot be treated as a single choice to make information visible or not; it must account for who gains access. 

Thus, the value of AI in this domain lies less in acting on users' behalf than in reducing the repeated labor around coordination while preserving users' authority over when and how support is enacted. Designers should therefore favor editable profiles, opt-in requests, and consent-based coordination mechanisms that give users control over role-differentiated visibility, allowing them to specify which information is shared with peers, supervisors, or the broader organization. Systems should also incorporate governance safeguards that limit how disclosed information can be accessed, retained, or repurposed, so that tools intended to support access do not become mechanisms for workplace surveillance or evaluation.

\textbf{Scaffold human judgment rather than replace it.}
CSCW research has long shown that collaborative systems can shape how people communicate, coordinate, and participate in group work \cite{samrose_meetingcoach_2021, samrose_coco_2018, mu_whispering_2024, kim_we_2023}. Our findings extend this perspective by showing that allyship involves not just coordination, but also ongoing interpersonal judgment about how to provide support in ways that are respectful, appropriate, and non-presumptive. Participants responded most positively to scenarios that helped them understand possible accommodations, surfaced relevant context, or suggested more reflective ways of responding, and were less comfortable with designs that replaced human expression or made socially consequential decisions on their own. This suggests that the role of AI in allyship-support tools is not to determine the ``correct'' social action, but to scaffold the process through which collaborators make better judgments themselves. In contrast to automation-oriented approaches that aim to reduce user involvement, allyship-support tools should preserve human responsibility for socially meaningful decisions while reducing the uncertainty surrounding those decisions. In practice, this means designing systems that suggest, prompt, or explain rather than systems that fully automate requests, offers, or decisions.

\textbf{Preserve authenticity and opportunities for interpersonal skill-building.}
Prior work on accessibility and collaboration has often focused on reducing burden and improving the efficiency of support practices \cite{mack_mixed_2021, wang_accessibility_2018, li_re-envisioning_2024}. While our findings align with these goals, they also surface a complementary concern: participants cared not only about whether support occurred, but about how it was enacted and what it signaled about the relationship. This extends prior work on AI-mediated social interaction, which has shown that AI can affect perceptions of social connection and authenticity in collaboration \cite{suh_ai_2021, johnson_exploring_2025, hwang_in_2024}. Designing for allyship, therefore, requires preserving opportunities for interpersonal engagement and skill-building. Rephrasing tools, reflection prompts, and communication aids may still be useful, but they should be framed as scaffolds for learning and self-reflection rather than substitutes for interpersonal work. In this sense, allyship-support tools should help people become better at navigating difficult support interactions over time, rather than simply performing those interactions for them or removing the productive friction \cite{romerolauro_resilience_2026, kapur_productive_2008} through which collaborators learn how to understand and support one another.

Together, these guidelines demonstrate that allyship-support tools should support accessibility as an ongoing collaborative practice. Allyship-support tools should help teams make needs actionable, reduce the labor of coordination, and support better judgment, while preserving the agency, authenticity, and trust through which inclusive collaboration is sustained. Rather than replacing the interpersonal work of allyship, the role of these tools is to better scaffold the conditions under which that work can happen.
\subsection{Considerations for AI in Allyship-Support Tools}
While the guidelines above outline how allyship-support tools can be designed, introducing AI with agentic capabilities into this role raises additional questions for future work. We began this project with the goal of designing an AI agent that could act as an ally. However, through our formative study and design process, we did not arrive at a fully autonomous agent. Instead, we developed a form of AI with agentic capabilities that supports allyship through text-based, private interactions and remains passive unless users have specified preferences in advance. In our study, this AI guided profile creation, mediated how requests were made, prompted teammates to act, and intervened in how support was communicated. A non-agentic version of these same ideas might take the form of a manual anonymous request option or user-invoked writing suggestions. By contrast, a more fully agentic AI might decide who should provide support or act on a teammate’s behalf. This distinction highlights that not all allyship-support workflows require AI; rather, AI’s distinctive value lies in cases where support depends on recognizing, timing, or initiating opportunities as collaboration unfolds, such as identifying when a stated access preference becomes relevant in a particular interaction or prompting teammates when support might otherwise be missed. The use of AI and the extent of its agentic capabilities raise several questions for future work, which we discuss below.

First, how should responsibility be structured when AI participates in coordination? Our findings show that responsibility for accessibility is already uneven and context-dependent, and can become diffuse in larger teams. When AI actively prompts teammates, recommends who should respond, or helps coordinate support, it does not simply make information available; it participates in assigning attention and action. Seeber et al. \cite{seeber2020machines} identify accountability as one of the central unresolved tensions in designing AI as a teammate, noting that when systems take on coordination roles previously held by humans, questions of who bears responsibility for outcomes become difficult to answer. In our context, this tension is heightened by the social stakes of accessibility: if an AI recommends the wrong person to respond, misjudges how urgent a need is, or prompts support in a way that unintentionally exposes a teammate, the result is not just inefficiency but potential harm to trust and inclusion.

As with many emerging technologies, especially those introduced into socially consequential settings, AI systems for allyship support would require oversight and verification rather than being trusted to coordinate support on their own. If AI prompts teammates to provide support, future systems would need ways to determine whether the support was actually offered, whether it addressed the person’s need, and whether the AI’s interpretation of the situation was appropriate. This may require human-in-the-loop mechanisms, such as allowing the person requesting support to confirm whether the need was resolved, enabling teammates to correct or reject AI-generated suggestions, or designating a human coordinator for unresolved requests. Thus, future work must examine how such systems can be monitored, verified, and corrected.

Second, when should AI intervene? Because an agentic AI can decide when to surface a need, when to prompt others, and how strongly to frame its recommendations, it risks being experienced as authoritative even when its read of the situation is incomplete. Flathmann et al. \cite{flathmann2024empirically} show that people often adapt their behavior to align with AI recommendations through social influence processes, even when those recommendations may not be optimal: a risk that is especially consequential in socially sensitive tasks such as allyship decisions. Zhang et al. \cite{zhang2023investigating} further found that an AI teammate's communication style significantly shapes who feels empowered to act. Shen et al. \cite{shen2024position} argue for a bidirectional approach to alignment: rather than designing AI systems that simply act on users' behalf, future systems should help people understand, critique, and verify AI outputs while supporting calibrated reliance and opportunities for human intervention. In our context, this might mean AI that surfaces its own uncertainty about when to intervene, or that invites teammates to calibrate how strongly it should prompt, thereby preserving human judgment while still reducing coordination burden.

Finally, what happens to relationships when AI takes on more of the interpersonal work involved in supporting allyship? AI that notices needs, suggests responses, or mediates communication may scaffold processes for allyship, but it may also reduce opportunities for teammates to practice the interpersonal judgment and direct communication through which allyship is developed. Fu et al. \cite{fu2024text} found that users of AI-mediated communication tools grew concerned about inauthenticity and overreliance, worrying that the tools would erode their personal voice and communicative agency. Prior work on AI in creative collaboration offers a parallel: Suh et al. \cite{suh_ai_2021} found that generative AI can act as social glue, helping collaborators establish common ground and reduce interpersonal friction, while in some cases shifting interaction toward evaluating AI outputs rather than deeply building on each other's ideas. In a higher-stakes context like accessibility, this consideration has additional consequences. Jung et al. \cite{jung_understanding_2026} and Das et al. \cite{das_that_2024} both show that accessibility in mixed-ability teams is accomplished through ongoing negotiation, not simply through information transfer or task completion. An agent that intervenes too readily may short-circuit exactly the relational processes through which teammates build shared understanding of each other's needs.

Taken together, these questions highlight that designing AI with agentic capabilities for allyship is not only about introducing new features, but about carefully defining the AI's role in collaboration. The challenge is to determine when the agent should guide, prompt, or coordinate, and when it should step back to preserve human judgment and interaction. The risks identified here are not unique to disability contexts, but they are intensified by the social and technical stakes of accessibility work. Future work should explore the role of AI specifically in allyship-support tools.

\subsection{Limitations and Future Work}
We used a scenario-based study, which, like any method, carries inherent limitations. While this approach is valuable for understanding participants’ perceptions of emerging design spaces before technologies are built and deployed, it cannot reveal how systems would actually be used in practice. As a result, our findings reflect participants’ perceptions, values, and concerns about using AI for allyship support, rather than how such tools would actually be used, negotiated, or resisted in ongoing collaboration. Because the scenarios were hypothetical, participants could only describe how they imagined they might respond, not how they would act in real-time team settings. As a complement, future work should therefore examine prototype or deployed systems in real mixed-ability teams to understand how AI designed to support allyship is used in practice, including how such tools build trust, redistribute access labor, support accountability, or introduce new coordination burdens over time.

Our scenarios focused on one part of the design space for AI-supported allyship by emphasizing text-based, private, and passive or consent-driven AI interactions. This focus was informed by our iterative design process and formative study. While this let us examine these design assumptions in depth, it also left other possibilities unexplored, such as anthropomorphic AI representations, more proactive AI behaviors, more public forms of support, and multimodal interactions. Participants’ reactions should therefore be interpreted in relation to the specific scenario configurations we presented, including what information the AI appeared to know, what it inferred from context, what it disclosed to teammates, and whether it sought consent before making a recommendation. Different configurations, such as more assertive recommendations, public prompts, or AI that acts without situational authorization, could be perceived as more intrusive or risky. Future work should explore opportunities and limitations of alternative design approaches.

We also note that our recruiting process likely favored participants with certain characteristics. We recruited participants with experience in mixed-ability teams and framed the study around allyship in virtual collaboration. As a result, participants may have been more comfortable discussing disability and accessibility, or more attentive to these issues, than the broader population. This may limit how widely our findings apply, especially for teams with less experience engaging with accessibility. Future work should examine perspectives from a wider range of participants, including those with less prior exposure to accessibility practices.

\section{Conclusion}
We conducted a scenario-based study with 18 participants (10 disabled and 8 non-disabled) to examine what AI designs can support allyship in virtual collaboration. Participants perceived AI as most useful when it helped teammates express, interpret, and coordinate around access needs, and less acceptable when it displaced user control, human judgment, or the relational meaning of support. Based on these findings, we proposed allyship-support tools that scaffold how teammates recognize, communicate, coordinate, and respond to access needs within ongoing collaborative relationships.


\bibliographystyle{ACM-Reference-Format}
\bibliography{references}

@book{saldana_2016,
  title={The Coding Manual for Qualitative Researchers},
  author={Salda{\~n}a, Johnny},
  year={2016},
  publisher={Sage}
}

@inproceedings{hwang_in_2024,
  author = {Hwang, Angel Hsing-Chi and Siy, John Oliver and Shelby, Renee and Lentz, Ashley},
  title = {In Whose Voice?: Examining AI Agent Representation of People in Social Interaction through Generative Speech},
  booktitle = {Proceedings of the 2024 ACM Designing Interactive Systems Conference},
  year = {2024},
  doi = {10.1145/3643834.3661555}
}

@inproceedings{mack_mixed_2021,
	address = {Virtual Event USA},
	title = {Mixed {Abilities} and {Varied} {Experiences}: a group autoethnography of a virtual summer internship},
	isbn = {978-1-4503-8306-6},
	shorttitle = {Mixed {Abilities} and {Varied} {Experiences}},
	url = {https://dl.acm.org/doi/10.1145/3441852.3471199},
	doi = {10.1145/3441852.3471199},
	language = {en},
	urldate = {2024-12-10},
	booktitle = {Proceedings of the 23rd {International} {ACM} {SIGACCESS} {Conference} on {Computers} and {Accessibility}},
	publisher = {ACM},
	author = {Mack, Kelly and Das, Maitraye and Jain, Dhruv and Bragg, Danielle and Tang, John and Begel, Andrew and Beneteau, Erin and Davis, Josh Urban and Glasser, Abraham and Park, Joon Sung and Potluri, Venkatesh},
	month = oct,
	year = {2021},
	pages = {1--13},
}

@inproceedings{alharbi_accessibility_2023,
	address = {Hamburg Germany},
	title = {Accessibility {Barriers}, {Conflicts}, and {Repairs}: {Understanding} the {Experience} of {Professionals} with {Disabilities} in {Hybrid} {Meetings}},
	isbn = {978-1-4503-9421-5},
	shorttitle = {Accessibility {Barriers}, {Conflicts}, and {Repairs}},
	url = {https://dl.acm.org/doi/10.1145/3544548.3581541},
	doi = {10.1145/3544548.3581541},
	language = {en},
	urldate = {2024-09-10},
	booktitle = {Proceedings of the 2023 {CHI} {Conference} on {Human} {Factors} in {Computing} {Systems}},
	publisher = {ACM},
	author = {Alharbi, Rahaf and Tang, John and Henderson, Karl},
	month = apr,
	year = {2023},
	pages = {1--15},
}

@article{tang2021understanding,
  title={Understanding the telework experience of people with disabilities},
  author={Tang, John},
  journal={Proceedings of the ACM on Human-Computer Interaction},
  volume={5},
  number={CSCW1},
  pages={1--27},
  year={2021},
  publisher={ACM New York, NY, USA}
}

@inproceedings{cha_you_2024,
	address = {Honolulu HI USA},
	title = {"{Do} {You} {Want} {Me} to {Participate} or {Not}?": {Investigating} the {Accessibility} of {Software} {Development} {Meetings} for {Blind} and {Low} {Vision} {Professionals}},
	isbn = {9798400703300},
	shorttitle = {"{Do} {You} {Want} {Me} to {Participate} or {Not}?},
	url = {https://dl.acm.org/doi/10.1145/3613904.3642130},
	doi = {10.1145/3613904.3642130},
	language = {en},
	urldate = {2024-09-10},
	booktitle = {Proceedings of the {CHI} {Conference} on {Human} {Factors} in {Computing} {Systems}},
	publisher = {ACM},
	author = {Cha, Yoonha and Figueira, Isabela and Ayala, Jessy and Edwards, Emory James and Garcia, Joshua and Van Der Hoek, André and Branham, Stacy Marie},
	month = may,
	year = {2024},
	pages = {1--17},
}

@article{siebold_disability_2025,
	title = {Disability {Meets} {Modality}: {A} {Sociotechnical} {Approach} to {Team} {Meetings}},
	volume = {9},
	shorttitle = {Disability {Meets} {Modality}},
	url = {https://dl.acm.org/doi/10.1145/3757580},
	doi = {10.1145/3757580},
	number = {7},
	urldate = {2026-02-27},
	journal = {Proc. ACM Hum.-Comput. Interact.},
	author = {Siebold, Connie and Paletz, Susannah B. F. and Reiter-Palmon, Roni and Lewis, Shevaun and Mont, Eusebia},
	month = oct,
	year = {2025},
	pages = {CSCW399:1--CSCW399:24},
}

@article{lacy_seat_2022,
	title = {A {Seat} at the {Virtual} {Table}: {Emergent} {Inclusion} in {Remote} {Meetings}},
	volume = {6},
	shorttitle = {A {Seat} at the {Virtual} {Table}},
	url = {https://dl.acm.org/doi/10.1145/3555151},
	doi = {10.1145/3555151},
	number = {CSCW2},
	urldate = {2026-02-27},
	journal = {Proc. ACM Hum.-Comput. Interact.},
	author = {Lacy, Amanda and Polsley, Seth and Ray, Samantha and Hammond, Tracy},
	month = nov,
	year = {2022},
	pages = {426:1--426:20},
}

@article{das_that_2024,
	title = {"{That} comes with a huge career cost:" {Understanding} {Collaborative} {Ideation} {Experiences} of {Disabled} {Professionals}},
	volume = {8},
	shorttitle = {"{That} comes with a huge career cost},
	url = {https://dl.acm.org/doi/10.1145/3641018},
	doi = {10.1145/3641018},
	number = {CSCW1},
	urldate = {2026-03-02},
	journal = {Proc. ACM Hum.-Comput. Interact.},
	author = {Das, Maitraye and Stangl, Abigale and Findlater, Leah},
	year = {2024},
	pages = {179:1--179:28},
}

@inproceedings{akter_beyond_2025,
	address = {New York, NY, USA},
	series = {{ASSETS} '25},
	title = {Beyond {Accessibility}: {Understanding} the {Ease} of {Use} and {Impacts} of {Digital} {Collaboration} {Tools} for {Blind} and {Low} {Vision} {Workers}},
	isbn = {979-8-4007-0676-9},
	shorttitle = {Beyond {Accessibility}},
	url = {https://dl.acm.org/doi/10.1145/3663547.3746332},
	doi = {10.1145/3663547.3746332},
	urldate = {2026-03-02},
	booktitle = {Proceedings of the 27th {International} {ACM} {SIGACCESS} {Conference} on {Computers} and {Accessibility}},
	publisher = {Association for Computing Machinery},
	author = {Akter, Taslima and Marathe, Aparajita S and Gergle, Darren and Piper, Anne Marie},
	month = oct,
	year = {2025},
	pages = {1--17},
}

@inproceedings{kushalnagar_teleconference_2020,
	address = {New York, NY, USA},
	series = {{ASSETS} '20},
	title = {Teleconference {Accessibility} and {Guidelines} for {Deaf} and {Hard} of {Hearing} {Users}},
	isbn = {978-1-4503-7103-2},
	url = {https://dl.acm.org/doi/10.1145/3373625.3417299},
	doi = {10.1145/3373625.3417299},
	urldate = {2026-03-02},
	booktitle = {Proceedings of the 22nd {International} {ACM} {SIGACCESS} {Conference} on {Computers} and {Accessibility}},
	publisher = {Association for Computing Machinery},
	author = {Kushalnagar, Raja S. and Vogler, Christian},
	month = oct,
	year = {2020},
	pages = {1--6},
}

@article{zolyomi_managing_2019,
	title = {Managing {Stress}: {The} {Needs} of {Autistic} {Adults} in {Video} {Calling}},
	volume = {3},
	shorttitle = {Managing {Stress}},
	url = {https://dl.acm.org/doi/10.1145/3359236},
	doi = {10.1145/3359236},
	number = {CSCW},
	urldate = {2026-03-02},
	journal = {Proc. ACM Hum.-Comput. Interact.},
	author = {Zolyomi, Annuska and Begel, Andrew and Waldern, Jennifer Frances and Tang, John and Barnett, Michael and Cutrell, Edward and McDuff, Daniel and Andrist, Sean and Morris, Meredith Ringel},
	month = nov,
	year = {2019},
	pages = {134:1--134:29},
}

@misc{otter_ai_2026,
  author       = {{Otter.ai}},
  title        = {Otter.ai: AI Meeting Notes, Transcription, and Summaries},
  year         = {2026},
  howpublished = {\url{https://otter.ai/}},
}

@book{miles2014qualitative,
  author    = {Miles, Matthew B. and Huberman, A. Michael and Salda{\~n}a, Johnny},
  title     = {Qualitative Data Analysis: A Methods Sourcebook},
  edition   = {3rd},
  publisher = {SAGE Publications},
  address   = {Thousand Oaks, CA},
  year      = {2014}
}

@misc{zoom_ai_captions_2025,
  author       = {{Zoom}},
  title        = {How AI is Helping Make Meetings More Accessible with Captions},
  year         = {2025},
  howpublished = {\url{https://www.zoom.com/en/blog/ai-accessible-meetings-captions/}},
}

@misc{google_workspace_ai_meetings_2026,
  author       = {{Google}},
  title        = {AI for Meetings and Video Conferencing in Google Workspace},
  year         = {2026},
  howpublished = {\url{https://workspace.google.com/resources/ai-for-meetings/}},
}

@misc{grammarly_2026,
  author       = {{Grammarly}},
  title        = {Grammarly Writing Assistant},
  year         = {2026},
  howpublished = {\url{https://app.grammarly.com/}},
}

@inproceedings{dangol2025want,
  title={“I Want to Think Like an SLP”: A Design Exploration of AI-Supported Home Practice in Speech Therapy},
  author={Dangol, Aayushi and Lewis, Aaleyah and Suh, Hyewon and Hong, Xuesi and Meadan, Hedda and Fogarty, James and Kientz, Julie A},
  booktitle={Proceedings of the 2025 CHI Conference on Human Factors in Computing Systems},
  pages={1--22},
  year={2025}
}

@article{zimmerman2017speed,
  title={Speed dating: Providing a menu of possible futures},
  author={Zimmerman, John and Forlizzi, Jodi},
  journal={She Ji: The Journal of Design, Economics, and Innovation},
  volume={3},
  number={1},
  pages={30--50},
  year={2017},
  publisher={Elsevier}
}

@article{seo2021impact,
  title={The impact of artificial intelligence on learner--instructor interaction in online learning},
  author={Seo, Kyoungwon and Tang, Joice and Roll, Ido and Fels, Sidney and Yoon, Dongwook},
  journal={International journal of educational technology in higher education},
  volume={18},
  number={1},
  pages={54},
  year={2021},
  publisher={Springer}
}

@inproceedings{luria2020social,
  title={Social boundaries for personal agents in the interpersonal space of the home},
  author={Luria, Michal and Zheng, Rebecca and Huffman, Bennett and Huang, Shuangni and Zimmerman, John and Forlizzi, Jodi},
  booktitle={Proceedings of the 2020 CHI conference on human factors in computing systems},
  pages={1--12},
  year={2020}
}

@inproceedings{chang2024dynamic,
  title={Dynamic Agent Affiliation: Who Should the AI Agent Work for in the Older Adult's Care Network?},
  author={Chang, Mai Lee and Lee, Alicia and Han, Nara and Huang, Anna and Sim{\~a}o, Hugo and Reig, Samantha and Mohammad Ali, Abdullah Ubed and Martinez, Rebekah and Khanuja, Neeta M and Zimmerman, John and others},
  booktitle={Proceedings of the 2024 ACM Designing Interactive Systems Conference},
  pages={1774--1788},
  year={2024}
}

@misc{washington_accessibility_etiquette,
  author       = {{Washington State Office of the Chief Information Officer}},
  title        = {Disability Etiquette for Respecting People with Disabilities},
  howpublished = {\url{https://wa.gov/accessibility-in-washington/accessibility-basics/disability-etiquette-for-respecting-people-with-disabilities}},
}

@misc{ndnrc_disability_etiquette,
  author       = {{National Disability Navigator Resource Collaborative}},
  title        = {What is Disability Etiquette?},
  howpublished = {\url{https://nationaldisabilitynavigator.org/ndnrc-materials/disability-guide/what-is-disability-etiquette/}},
}

@misc{cohen_disability_etiquette_epva,
  author       = {Judy Cohen},
  title        = {Disability Etiquette: Tips on Interacting with People with Disabilities},
  institution  = {Eastern Paralyzed Veterans Association},
  howpublished = {\url{https://ds.gmu.edu/wp-content/uploads/2020/08/disability-etiquette-epva.pdf}},
}

@misc{unitedspinal_top10_etiquette,
  author       = {{United Spinal Association}},
  title        = {Top 10 Disability Etiquette Tips},
  howpublished = {\url{https://unitedspinal.org/top-10-disability-etiquette/}},
}

@misc{disabilityin_etiquette_guide,
  author       = {{Disability:IN}},
  title        = {Disability Etiquette: A Starting Guide},
  howpublished = {\url{https://disabilityin.org/resource/disability-etiquette---a-starting-guide}},
}

@inproceedings{liao2023proactive,
  title={Proactive conversational agents in the post-chatgpt world},
  author={Liao, Lizi and Yang, Grace Hui and Shah, Chirag},
  booktitle={Proceedings of the 46th international ACM SIGIR conference on research and development in information retrieval},
  pages={3452--3455},
  year={2023}
}

@book{norman2009design,
  title={The design of future things},
  author={Norman, Don},
  year={2009},
  publisher={Basic books}
}

@article{meurisch2020exploring,
  title={Exploring user expectations of proactive AI systems},
  author={Meurisch, Christian and Mihale-Wilson, Cristina A and Hawlitschek, Adrian and Giger, Florian and M{\"u}ller, Florian and Hinz, Oliver and M{\"u}hlh{\"a}user, Max},
  journal={Proceedings of the ACM on Interactive, Mobile, Wearable and Ubiquitous Technologies},
  volume={4},
  number={4},
  pages={1--22},
  year={2020},
  publisher={ACM New York, NY, USA}
}

@article{nowak_biocca,
author = {Nowak, Kristine L. and Biocca, Frank},
title = {The effect of the agency and anthropomorphism of users' sense of telepresence, copresence, and social presence in virtual environments},
year = {2003},
issue_date = {October 2003},
publisher = {MIT Press},
address = {Cambridge, MA, USA},
volume = {12},
number = {5},
issn = {1054-7460},
url = {https://doi-org.proxy.library.cornell.edu/10.1162/105474603322761289},
doi = {10.1162/105474603322761289},
journal = {Presence: Teleoper. Virtual Environ.},
month = oct,
pages = {481–494},
numpages = {14}
}

@inproceedings{sauppe2015social,
  title={The social impact of a robot co-worker in industrial settings},
  author={Saupp{\'e}, Allison and Mutlu, Bilge},
  booktitle={Proceedings of the 33rd annual ACM conference on human factors in computing systems},
  pages={3613--3622},
  year={2015}
}

@inproceedings{ehsan2021expanding,
  title={Expanding explainability: Towards social transparency in ai systems},
  author={Ehsan, Upol and Liao, Q Vera and Muller, Michael and Riedl, Mark O and Weisz, Justin D},
  booktitle={Proceedings of the 2021 CHI conference on human factors in computing systems},
  pages={1--19},
  year={2021}
}

@inproceedings{chan2024visibility,
  title={Visibility into AI agents},
  author={Chan, Alan and Ezell, Carson and Kaufmann, Max and Wei, Kevin and Hammond, Lewis and Bradley, Herbie and Bluemke, Emma and Rajkumar, Nitarshan and Krueger, David and Kolt, Noam and others},
  booktitle={Proceedings of the 2024 ACM conference on fairness, accountability, and transparency},
  pages={958--973},
  year={2024}
}

@article{das_towards_2021,
	title = {Towards {Accessible} {Remote} {Work}: {Understanding} {Work}-from-{Home} {Practices} of {Neurodivergent} {Professionals}},
	volume = {5},
	shorttitle = {Towards {Accessible} {Remote} {Work}},
	url = {https://dl.acm.org/doi/10.1145/3449282},
	doi = {10.1145/3449282},
	number = {CSCW1},
	urldate = {2026-03-02},
	journal = {Proc. ACM Hum.-Comput. Interact.},
	author = {Das, Maitraye and Tang, John and Ringland, Kathryn E. and Piper, Anne Marie},
	month = apr,
	year = {2021},
	pages = {183:1--183:30},
}

@inproceedings{jung_understanding_2026,
  author    = {Jung, Crescentia and Cheng, Kexin and Heung, Sharon and Jung, Malte F. and Azenkot, Shiri},
  title     = {Understanding How Accessibility Practices Impact Teamwork in Mixed-Ability Teams that Collaborate Virtually},
  booktitle = {Proceedings of the 2026 CHI Conference on Human Factors in Computing Systems (CHI '26)},
  publisher = {Association for Computing Machinery},
  address   = {New York, NY, USA},
  year      = {2026},
  doi       = {10.1145/3772318.3790419}
}

@article{lindsay_systematic_2018,
	title = {A systematic review of workplace disclosure and accommodation requests among youth and young adults with disabilities},
	volume = {40},
	issn = {0963-8288},
	url = {https://doi.org/10.1080/09638288.2017.1363824},
	doi = {10.1080/09638288.2017.1363824},
	number = {25},
	urldate = {2026-03-04},
	journal = {Disability and Rehabilitation},
	publisher = {Taylor \& Francis},
	author = {Lindsay, Sally and Cagliostro, Elaine and Carafa, Gabriella},
	month = dec,
	year = {2018},
	pages = {2971--2986},
}

@article{von_schrader_perspectives_2014,
	title = {Perspectives on {Disability} {Disclosure}: {The} {Importance} of {Employer} {Practices} and {Workplace} {Climate}},
	volume = {26},
	issn = {1573-3378},
	shorttitle = {Perspectives on {Disability} {Disclosure}},
	url = {https://doi.org/10.1007/s10672-013-9227-9},
	doi = {10.1007/s10672-013-9227-9},
	language = {en},
	number = {4},
	urldate = {2026-03-04},
	journal = {Employee Responsibilities and Rights Journal},
	author = {von Schrader, Sarah and Malzer, Valerie and Bruyère, Susanne},
	month = dec,
	year = {2014},
	pages = {237--255},
}

@inproceedings{samrose_meetingcoach_2021,
	address = {New York, NY, USA},
	series = {{CHI} '21},
	title = {{MeetingCoach}: {An} {Intelligent} {Dashboard} for {Supporting} {Effective} \& {Inclusive} {Meetings}},
	isbn = {978-1-4503-8096-6},
	shorttitle = {{MeetingCoach}},
	url = {https://dl.acm.org/doi/10.1145/3411764.3445615},
	doi = {10.1145/3411764.3445615},
	urldate = {2026-03-04},
	booktitle = {Proceedings of the 2021 {CHI} {Conference} on {Human} {Factors} in {Computing} {Systems}},
	publisher = {Association for Computing Machinery},
	author = {Samrose, Samiha and McDuff, Daniel and Sim, Robert and Suh, Jina and Rowan, Kael and Hernandez, Javier and Rintel, Sean and Moynihan, Kevin and Czerwinski, Mary},
	month = may,
	year = {2021},
	pages = {1--13},
}

@article{samrose_coco_2018,
	title = {{CoCo}: {Collaboration} {Coach} for {Understanding} {Team} {Dynamics} during {Video} {Conferencing}},
	volume = {1},
	shorttitle = {{CoCo}},
	url = {https://dl.acm.org/doi/10.1145/3161186},
	doi = {10.1145/3161186},
	number = {4},
	urldate = {2026-03-04},
	journal = {Proc. ACM Interact. Mob. Wearable Ubiquitous Technol.},
	author = {Samrose, Samiha and Zhao, Ru and White, Jeffery and Li, Vivian and Nova, Luis and Lu, Yichen and Ali, Mohammad Rafayet and Hoque, Mohammed Ehsan},
	month = jan,
	year = {2018},
	pages = {160:1--160:24},
}

@inproceedings{li_re-envisioning_2024,
	address = {New York, NY, USA},
	series = {{DIS} '24},
	title = {Re-envisioning {Remote} {Meetings}: {Co}-designing {Inclusive} and {Empowering} {Videoconferencing} with {People} {Who} {Stutter}},
	isbn = {979-8-4007-0583-0},
	shorttitle = {Re-envisioning {Remote} {Meetings}},
	url = {https://dl.acm.org/doi/10.1145/3643834.3661533},
	doi = {10.1145/3643834.3661533},
	urldate = {2026-03-04},
	booktitle = {Proceedings of the 2024 {ACM} {Designing} {Interactive} {Systems} {Conference}},
	publisher = {Association for Computing Machinery},
	author = {Li, Jingjin and Wu, Shaomei and Leshed, Gilly},
	month = jul,
	year = {2024},
	pages = {1926--1941},
}

@inproceedings{kim_we_2023,
	address = {New York, NY, USA},
	series = {{CHI} '23},
	title = {"{We} {Speak} {Visually}": {User}-{Generated} {Icons} for {Better} {Video}-{Mediated} {Mixed}-{Group} {Communications} {Between} {Deaf} and {Hearing} {Participants}},
	isbn = {978-1-4503-9421-5},
	shorttitle = {"{We} {Speak} {Visually}"},
	url = {https://dl.acm.org/doi/10.1145/3544548.3581151},
	doi = {10.1145/3544548.3581151},
	urldate = {2026-03-04},
	booktitle = {Proceedings of the 2023 {CHI} {Conference} on {Human} {Factors} in {Computing} {Systems}},
	publisher = {Association for Computing Machinery},
	author = {Kim, Yeon Soo and Im, Hyeonjeong and Lee, Sunok and Cho, Haena and Lee, Sangsu},
	month = apr,
	year = {2023},
	pages = {1--16},
}

@article{lee_supporting_2025,
	title = {Supporting from afar: {Exploring} {Practices} and {Challenges} of {Remote} {Support}-{Giving} for {Mental} {Health}},
	volume = {9},
	shorttitle = {Supporting from afar},
	url = {https://dl.acm.org/doi/10.1145/3757662},
	doi = {10.1145/3757662},
	number = {7},
	urldate = {2026-03-04},
	journal = {Proc. ACM Hum.-Comput. Interact.},
	author = {Lee, Tuan-He and Nambiar, Keertana and Wong, Hoi Ki Jenny and Leshed, Gilly},
	month = oct,
	year = {2025},
	pages = {CSCW481:1--CSCW481:26},
}

@article{wang_practice_2025,
	title = {The {Practice} of {Online} {Peer} {Counseling} and the {Potential} for {AI}-{Powered} {Support} {Tools}},
	volume = {9},
	url = {https://dl.acm.org/doi/10.1145/3711089},
	doi = {10.1145/3711089},
	number = {2},
	urldate = {2026-03-04},
	journal = {Proc. ACM Hum.-Comput. Interact.},
	author = {Wang, Tony and Bruckman, Amy S and Yang, Diyi},
	month = may,
	year = {2025},
	pages = {CSCW191:1--CSCW191:33},
}

@inproceedings{lasecki2013answering,
  title={Answering visual questions with conversational crowd assistants},
  author={Lasecki, Walter S and Thiha, Phyo and Zhong, Yu and Brady, Erin and Bigham, Jeffrey P},
  booktitle={Proceedings of the 15th International ACM SIGACCESS Conference on Computers and Accessibility},
  pages={1--8},
  year={2013}
}

@inproceedings{brady2013investigating,
  title={Investigating the appropriateness of social network question asking as a resource for blind users},
  author={Brady, Erin L and Zhong, Yu and Morris, Meredith Ringel and Bigham, Jeffrey P},
  booktitle={Proceedings of the 2013 conference on Computer supported cooperative work},
  pages={1225--1236},
  year={2013}
}

@inproceedings{brady2013visual,
  title={Visual challenges in the everyday lives of blind people},
  author={Brady, Erin and Morris, Meredith Ringel and Zhong, Yu and White, Samuel and Bigham, Jeffrey P},
  booktitle={Proceedings of the SIGCHI conference on human factors in computing systems},
  pages={2117--2126},
  year={2013}
}

@inproceedings{brady2014friendsourcing,
  title={Friendsourcing for the greater good: Perceptions of social microvolunteering},
  author={Brady, Erin and Morris, Meredith and Bigham, Jeffrey},
  booktitle={Proceedings of the AAAI Conference on Human Computation and Crowdsourcing},
  volume={2},
  pages={6--7},
  year={2014}
}

@article{brady2015using,
  title={Using social microvolunteering to answer visual questions from blind users},
  author={Brady, Erin},
  journal={ACM SIGACCESS Accessibility and Computing},
  number={111},
  pages={26--29},
  year={2015},
  publisher={ACM New York, NY, USA}
}

@inproceedings{guo2016vizlens,
  title={Vizlens: A robust and interactive screen reader for interfaces in the real world},
  author={Guo, Anhong and Chen, Xiang'Anthony' and Qi, Haoran and White, Samuel and Ghosh, Suman and Asakawa, Chieko and Bigham, Jeffrey P},
  booktitle={Proceedings of the 29th annual symposium on user interface software and technology},
  pages={651--664},
  year={2016}
}

@inproceedings{gleason2016vizmap,
  title={VizMap: Accessible visual information through crowdsourced map reconstruction},
  author={Gleason, Cole and Guo, Anhong and Laput, Gierad and Kitani, Kris and Bigham, Jeffrey P},
  booktitle={Proceedings of the 18th International ACM SIGACCESS Conference on Computers and Accessibility},
  pages={273--274},
  year={2016}
}

@inproceedings{burton2012crowdsourcing,
  title={Crowdsourcing subjective fashion advice using VizWiz: challenges and opportunities},
  author={Burton, Michele A and Brady, Erin and Brewer, Robin and Neylan, Callie and Bigham, Jeffrey P and Hurst, Amy},
  booktitle={Proceedings of the 14th international ACM SIGACCESS conference on Computers and accessibility},
  pages={135--142},
  year={2012}
}

@inproceedings{bigham2010vizwiz,
  title={Vizwiz: nearly real-time answers to visual questions},
  author={Bigham, Jeffrey P and Jayant, Chandrika and Ji, Hanjie and Little, Greg and Miller, Andrew and Miller, Robert C and Miller, Robin and Tatarowicz, Aubrey and White, Brandyn and White, Samual and others},
  booktitle={Proceedings of the 23nd annual ACM symposium on User interface software and technology},
  pages={333--342},
  year={2010}
}

@article{wobbrock_ability-based_2011,
  author = {Wobbrock, Jacob O. and Kane, Shaun K. and Gajos, Krzysztof Z. and Harada, Susumu and Froehlich, Jon},
  title = {Ability-Based Design: Concept, Principles and Examples},
  journal = {ACM Transactions on Accessible Computing},
  volume = {3},
  number = {3},
  pages = {9:1--9:27},
  year = {2011},
  doi = {10.1145/1952383.1952384}
}

@article{flathmann2024empirically,
  title={Empirically understanding the potential impacts and process of social influence in human-AI teams},
  author={Flathmann, Christopher and Duan, Wen and Mcneese, Nathan J and Hauptman, Allyson and Zhang, Rui},
  journal={Proceedings of the ACM on Human-Computer Interaction},
  volume={8},
  number={CSCW1},
  pages={1--32},
  year={2024},
  publisher={ACM New York, NY, USA}
}

@article{zhang2023investigating,
  title={Investigating AI teammate communication strategies and their impact in human-AI teams for effective teamwork},
  author={Zhang, Rui and Duan, Wen and Flathmann, Christopher and McNeese, Nathan and Freeman, Guo and Williams, Alyssa},
  journal={Proceedings of the ACM on Human-Computer Interaction},
  volume={7},
  number={CSCW2},
  pages={1--31},
  year={2023},
  publisher={ACM New York, NY, USA}
}

@article{seeber2020machines,
  title={Machines as teammates: A research agenda on AI in team collaboration},
  author={Seeber, Isabella and Bittner, Eva and Briggs, Robert O and De Vreede, Triparna and De Vreede, Gert-Jan and Elkins, Aaron and Maier, Ronald and Merz, Alexander B and Oeste-Rei{\ss}, Sarah and Randrup, Nils and others},
  journal={Information \& management},
  volume={57},
  number={2},
  pages={103174},
  year={2020},
  publisher={Elsevier}
}

@article{shen2024position,
  title={Position: Towards Bidirectional Human-AI Alignment},
  author={Shen, Hua and Knearem, Tiffany and Ghosh, Reshmi and Alkiek, Kenan and Krishna, Kundan and Liu, Yachuan and Ma, Ziqiao and Petridis, Savvas and Peng, Yi-Hao and Qiwei, Li and others},
  journal={arXiv preprint arXiv:2406.09264},
  year={2024}
}

@inproceedings{fu2024text,
  title={From text to self: Users’ perception of AIMC tools on interpersonal communication and self},
  author={Fu, Yue and Foell, Sami and Xu, Xuhai and Hiniker, Alexis},
  booktitle={Proceedings of the 2024 CHI Conference on Human Factors in Computing Systems},
  pages={1--17},
  year={2024}
}

@inproceedings{suh_ai_2021,
	address = {New York, NY, USA},
	series = {{CHI} '21},
	title = {{AI} as {Social} {Glue}: {Uncovering} the {Roles} of {Deep} {Generative} {AI} during {Social} {Music} {Composition}},
	isbn = {978-1-4503-8096-6},
	shorttitle = {{AI} as {Social} {Glue}},
	url = {https://dl.acm.org/doi/10.1145/3411764.3445219},
	doi = {10.1145/3411764.3445219},
	urldate = {2026-03-04},
	booktitle = {Proceedings of the 2021 {CHI} {Conference} on {Human} {Factors} in {Computing} {Systems}},
	publisher = {Association for Computing Machinery},
	author = {Suh, Minhyang (Mia) and Youngblom, Emily and Terry, Michael and Cai, Carrie J},
	month = may,
	year = {2021},
	pages = {1--11},
}

@inproceedings{carroll_scenario-based_2026,
author = {Carroll, John M. and Jo, Jeongwon and Kim,Jiyoon and Lin, Ya-Fang and Chen, Ester},
title = {What Happened to Scenario-Based Design in HCI?: A Scoping Review},
year = {2026},
isbn = {9798400722783},
publisher = {Association for Computing Machinery},
address = {New York, NY, USA},
url = {https://doi-org.proxy.library.cornell.edu/10.1145/3772318.3790899},
doi = {10.1145/3772318.3790899},
booktitle = {Proceedings of the 2026 CHI Conference on Human Factors in Computing Systems},
articleno = {1019},
numpages = {29},
location = {
},
series = {CHI '26}
}

@article{johnson_exploring_2025,
	title = {Exploring {Collaborative} {GenAI} {Agents} in {Synchronous} {Group} {Settings}: {Eliciting} {Team} {Perceptions} and {Design} {Considerations} for the {Future} of {Work}},
	volume = {9},
	shorttitle = {Exploring {Collaborative} {GenAI} {Agents} in {Synchronous} {Group} {Settings}},
	url = {https://dl.acm.org/doi/10.1145/3757595},
	doi = {10.1145/3757595},
	number = {7},
	urldate = {2026-03-04},
	journal = {Proc. ACM Hum.-Comput. Interact.},
	author = {Johnson, Janet G. and Peralta, Macarena and Kaur, Mansanjam and Huang, Ruijie Sophia and Zhao, Sheng and Guan, Ruijia and Rajaram, Shwetha and Nebeling, Michael},
	month = oct,
	year = {2025},
	pages = {CSCW414:1--CSCW414:33},
}

@article{pandey_understanding_2021,
	title = {Understanding {Accessibility} and {Collaboration} in {Programming} for {People} with {Visual} {Impairments}},
	volume = {5},
	url = {https://dl.acm.org/doi/10.1145/3449203},
	doi = {10.1145/3449203},
	number = {CSCW1},
	urldate = {2026-03-04},
	journal = {Proc. ACM Hum.-Comput. Interact.},
	author = {Pandey, Maulishree and Kameswaran, Vaishnav and Rao, Hrishikesh V. and O'Modhrain, Sile and Oney, Steve},
	month = apr,
	year = {2021},
	pages = {129:1--129:30},
}

@inproceedings{yildiz_virtual_2023,
	address = {New York, NY, USA},
	series = {{CHI} '23},
	title = {Virtual {Collaboration} {Tools} for {Mixed}-{Ability} {Workspaces}: {A} {Cross} {Disability} {Solidarity} {Case} from {Turkey}},
	isbn = {978-1-4503-9421-5},
	shorttitle = {Virtual {Collaboration} {Tools} for {Mixed}-{Ability} {Workspaces}},
	url = {https://dl.acm.org/doi/10.1145/3544548.3580937},
	doi = {10.1145/3544548.3580937},
	urldate = {2026-03-03},
	booktitle = {Proceedings of the 2023 {CHI} {Conference} on {Human} {Factors} in {Computing} {Systems}},
	publisher = {Association for Computing Machinery},
	author = {Yildiz, Zeynep and Subasi, Ozge},
	month = apr,
	year = {2023},
	pages = {1--11},
}

@article{wang_accessibility_2018,
	title = {Accessibility in {Action}: {Co}-{Located} {Collaboration} among {Deaf} and {Hearing} {Professionals}},
	volume = {2},
	shorttitle = {Accessibility in {Action}},
	url = {https://dl.acm.org/doi/10.1145/3274449},
	doi = {10.1145/3274449},
	number = {CSCW},
	urldate = {2026-03-04},
	journal = {Proc. ACM Hum.-Comput. Interact.},
	author = {Wang, Emily Q. and Piper, Anne Marie},
	month = nov,
	year = {2018},
	pages = {180:1--180:25},
}

@article{brown_what_2013,
	title = {What does it mean to be an ally?: {The} perception of allies from the perspective of people of color},
	volume = {43},
	copyright = {© 2013 Wiley Periodicals, Inc},
	issn = {1559-1816},
	shorttitle = {What does it mean to be an ally?},
	url = {https://onlinelibrary.wiley.com/doi/abs/10.1111/jasp.12172},
	doi = {10.1111/jasp.12172},
	language = {en},
	number = {11},
	urldate = {2026-03-05},
	journal = {Journal of Applied Social Psychology},
	author = {Brown, Kendrick T. and Ostrove, Joan M.},
	year = {2013},
	pages = {2211--2222},
}

@article{mcdonnell_social_2021,
	title = {Social, {Environmental}, and {Technical}: {Factors} at {Play} in the {Current} {Use} and {Future} {Design} of {Small}-{Group} {Captioning}},
	volume = {5},
	issn = {2573-0142},
	shorttitle = {Social, {Environmental}, and {Technical}},
	url = {https://dl.acm.org/doi/10.1145/3479578},
	doi = {10.1145/3479578},
	language = {en},
	number = {CSCW2},
	urldate = {2024-09-10},
	journal = {Proceedings of the ACM on Human-Computer Interaction},
	author = {McDonnell, Emma J. and Liu, Ping and Goodman, Steven M. and Kushalnagar, Raja and Froehlich, Jon E. and Findlater, Leah},
	month = oct,
	year = {2021},
	pages = {1--25},
}

@article{ostrove_one_2006,
	title = {"{One} lady was so busy staring at me she walked into a wall": {Interability} {Relations} from the {Perspective} of {Women} with {Disabilities}},
	volume = {26},
	issn = {2159-8371},
	shorttitle = {"{One} lady was so busy staring at me she walked into a wall"},
	url = {https://dsq-sds.org/article/id/5510/},
	doi = {10.18061/dsq.v26i3.717},
	language = {None},
	number = {3},
	urldate = {2026-03-05},
	journal = {Disability Studies Quarterly},
	publisher = {The Ohio State University Libraries},
	author = {Ostrove, Joan M. and Crawford, Danette},
	month = jun,
	year = {2006},
}

@inproceedings{bennett2018interdependence,
  title={Interdependence as a frame for assistive technology research and design},
  author={Bennett, Cynthia L and Brady, Erin and Branham, Stacy M},
  booktitle={Proceedings of the 20th international acm sigaccess conference on computers and accessibility},
  pages={161--173},
  year={2018}
}

@inproceedings{bennett_care_2020,
	address = {New York, NY, USA},
	series = {{CHI} '20},
	title = {The {Care} {Work} of {Access}},
	isbn = {978-1-4503-6708-0},
	url = {https://dl.acm.org/doi/10.1145/3313831.3376568},
	doi = {10.1145/3313831.3376568},
	urldate = {2026-03-05},
	booktitle = {Proceedings of the 2020 {CHI} {Conference} on {Human} {Factors} in {Computing} {Systems}},
	publisher = {Association for Computing Machinery},
	author = {Bennett, Cynthia L. and Rosner, Daniela K. and Taylor, Alex S.},
	month = apr,
	year = {2020},
	pages = {1--15},
}

@inproceedings{branham_collaborative_2015,
	address = {New York, NY, USA},
	series = {{CHI} '15},
	title = {Collaborative {Accessibility}: {How} {Blind} and {Sighted} {Companions} {Co}-{Create} {Accessible} {Home} {Spaces}},
	isbn = {978-1-4503-3145-6},
	shorttitle = {Collaborative {Accessibility}},
	url = {https://doi.org/10.1145/2702123.2702511},
	doi = {10.1145/2702123.2702511},
	urldate = {2026-03-05},
	booktitle = {Proceedings of the 33rd {Annual} {ACM} {Conference} on {Human} {Factors} in {Computing} {Systems}},
	publisher = {Association for Computing Machinery},
	author = {Branham, Stacy M. and Kane, Shaun K.},
	month = apr,
	year = {2015},
	pages = {2373--2382},
}

@inproceedings{mu_whispering_2024,
	address = {New York, NY, USA},
	series = {{CHI} '24},
	title = {Whispering {Through} {Walls}: {Towards} {Inclusive} {Backchannel} {Communication} in {Hybrid} {Meetings}},
	isbn = {979-8-4007-0330-0},
	shorttitle = {Whispering {Through} {Walls}},
	url = {https://dl.acm.org/doi/10.1145/3613904.3642419},
	doi = {10.1145/3613904.3642419},
	urldate = {2026-03-05},
	booktitle = {Proceedings of the 2024 {CHI} {Conference} on {Human} {Factors} in {Computing} {Systems}},
	publisher = {Association for Computing Machinery},
	author = {Mu, Qianqian and Borowski, Marcel and Grønbæk, Jens Emil Sloth and Bødker, Susanne and Hoggan, Eve},
	month = may,
	year = {2024},
	pages = {1--16},
}

@article{ostrove_reflections_2009,
	title = {Reflections on the {K}-12 years in public schools: {Relations} with hearing teachers and peers from the perspective of deaf and hard-of-hearing adults},
	volume = {29},
	issn = {2159-8371},
	shorttitle = {Reflections on the {K}-12 years in public schools},
	url = {https://dsq-sds.org/article/id/498/},
	doi = {10.18061/dsq.v29i3.931},
	language = {None},
	number = {3},
	urldate = {2026-03-05},
	journal = {Disability Studies Quarterly},
	publisher = {The Ohio State University Libraries},
	author = {Ostrove, Joan M. and Oliva, Gina A. and Katowitz, Abigail},
	month = jul,
	year = {2009},
}

@article{mackay1999paper,
  author    = {Mackay, Wendy E.},
  title     = {Is Paper Safer? The Role of Paper Flight Strips in Air Traffic Control},
  journal   = {ACM Transactions on Computer-Human Interaction},
  volume    = {6},
  number    = {4},
  pages     = {311--340},
  year      = {1999},
  publisher = {ACM},
  doi       = {10.1145/331490.331491}
}

@inproceedings{lindtner2016reconstituting,
  author    = {Lindtner, Silvia and Bardzell, Shaowen and Bardzell, Jeffrey},
  title     = {Reconstituting the Utopian Vision of Making: HCI Ninja from the Ground Up},
  booktitle = {Proceedings of the 2016 CHI Conference on Human Factors in Computing Systems},
  series    = {CHI '16},
  year      = {2016},
  publisher = {ACM},
  address   = {New York, NY, USA},
  doi       = {10.1145/2858036.2858506}
}

@misc{buch_workwithme_2020,
author = {Buch, Bastian},
title = {The ``Work with me User Guide''---An Extremely Simple Yet Powerful Tool to Align with Your Team},
year = {2020},
month = may,
day = {25},
howpublished = {Medium},
url = {https://medium.com/@bastianbuch/the-work-with-me-user-guide-an-extremely-simple-yet-powerful-tool-to-align-with-your-team-f5d3e1b42339}
}

@misc{scrum_team_working_agreement_nd,
author = {{Scrum.org}},
title = {Creating a Team Working Agreement},
howpublished = {Scrum.org Resources},
url = {https://www.scrum.org/resources/creating-team-working-agreement}
}

@misc{atlassian_user_manual_nd,
author = {{Atlassian}},
title = {How to Create a Personal User Manual for Work},
howpublished = {Atlassian Team Playbook},
url = {https://www.atlassian.com/team-playbook/plays/my-user-manual}
}

@misc{maymin_howtoworkwithme_2020,
author = {Maymin, Senia},
title = {The ``How to Work with Me'' Document},
year = {2020},
month = jul,
day = {21},
howpublished = {LinkedIn},
url = {https://www.linkedin.com/pulse/how-work-me-document-senia-maymin/}
}

@inproceedings{romerolauro_resilience_2026,
author = {Romero Lauro, Quentin and Gautam, Aakash and Kotturi, Yasmine},
title = {Towards Designing for Resilience: Community-Centered Deployment of an AI Business Planning Tool in a Pittsburgh Small Business Center},
booktitle = {Proceedings of the 2026 CHI Conference on Human Factors in Computing Systems},
series = {CHI '26},
year = {2026},
publisher = {Association for Computing Machinery},
address = {New York, NY, USA},
articleno = {669},
pages = {1--19},
numpages = {19},
doi = {10.1145/3772318.3791654},
url = {https://doi.org/10.1145/3772318.3791654}
}

@article{kapur_productive_2008,
author = {Kapur, Manu},
title = {Productive Failure},
journal = {Cognition and Instruction},
volume = {26},
number = {3},
pages = {379--424},
year = {2008},
publisher = {Taylor & Francis},
doi = {10.1080/07370000802212669},
url = {https://doi.org/10.1080/07370000802212669}
}

@inproceedings{zhao2026accessibility,
  author    = {Zhao, Yichun and Nacenta, Miguel A. and Sukhai, Mahadeo A. and Somanath, Sowmya},
  title     = {Accessibility-Driven Information Transformations in Mixed-Visual Ability Work Teams},
  booktitle = {Proceedings of the 2026 CHI Conference on Human Factors in Computing Systems},
  series    = {CHI '26},
  year      = {2026},
  pages     = {1--14},
  publisher = {ACM},
  address   = {New York, NY, USA},
  doi       = {10.1145/3772318.3790872},
  isbn      = {979-8-4007-2278-3/2026/04}
}

\appendix
\clearpage
\begin{figure*}[!t]
    \section{Storyboards of Scenarios}
    \label{app:storyboards}
    \centering
    \includegraphics[width=0.6\textwidth]{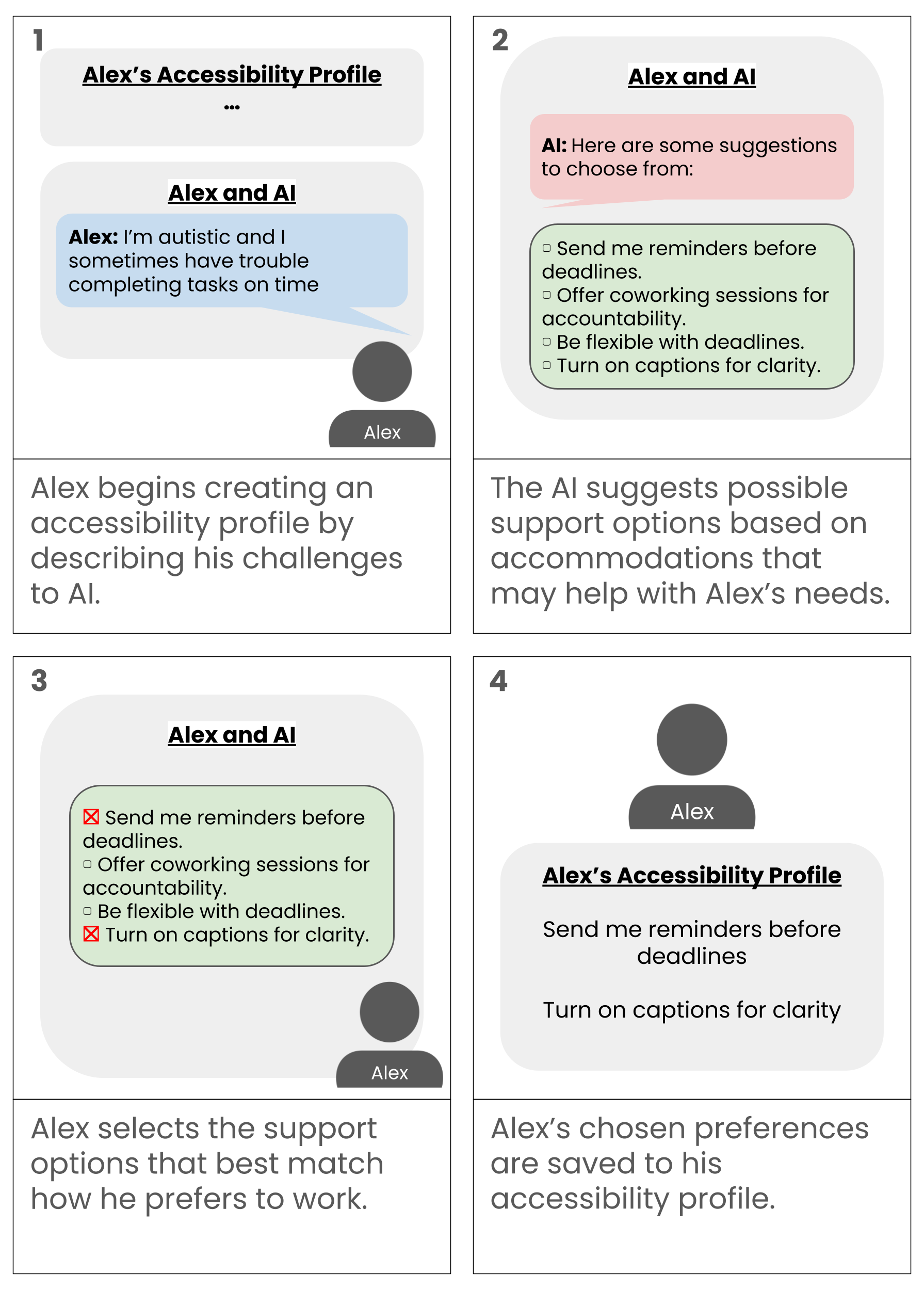}
    \caption{Scenario 1: Accessibility Profile Support.}
    \Description{Four-panel storyboard with numbered panels, profile and chat interface elements, colored message bubbles, checklist-style options, and captions beneath each panel. Alex tells the AI that he is autistic and sometimes has trouble completing tasks on time. The AI suggests possible support options, Alex selects reminders and captions, and those choices are saved to his accessibility profile. The panel captions describe the sequence from describing a challenge, to reviewing suggested supports, to selecting preferred options, to saving those preferences in the profile.}
    \label{fig:appendix_scenario1}
\end{figure*}

\begin{figure*}[htbp]
    \centering
    \includegraphics[width=0.85\textwidth]{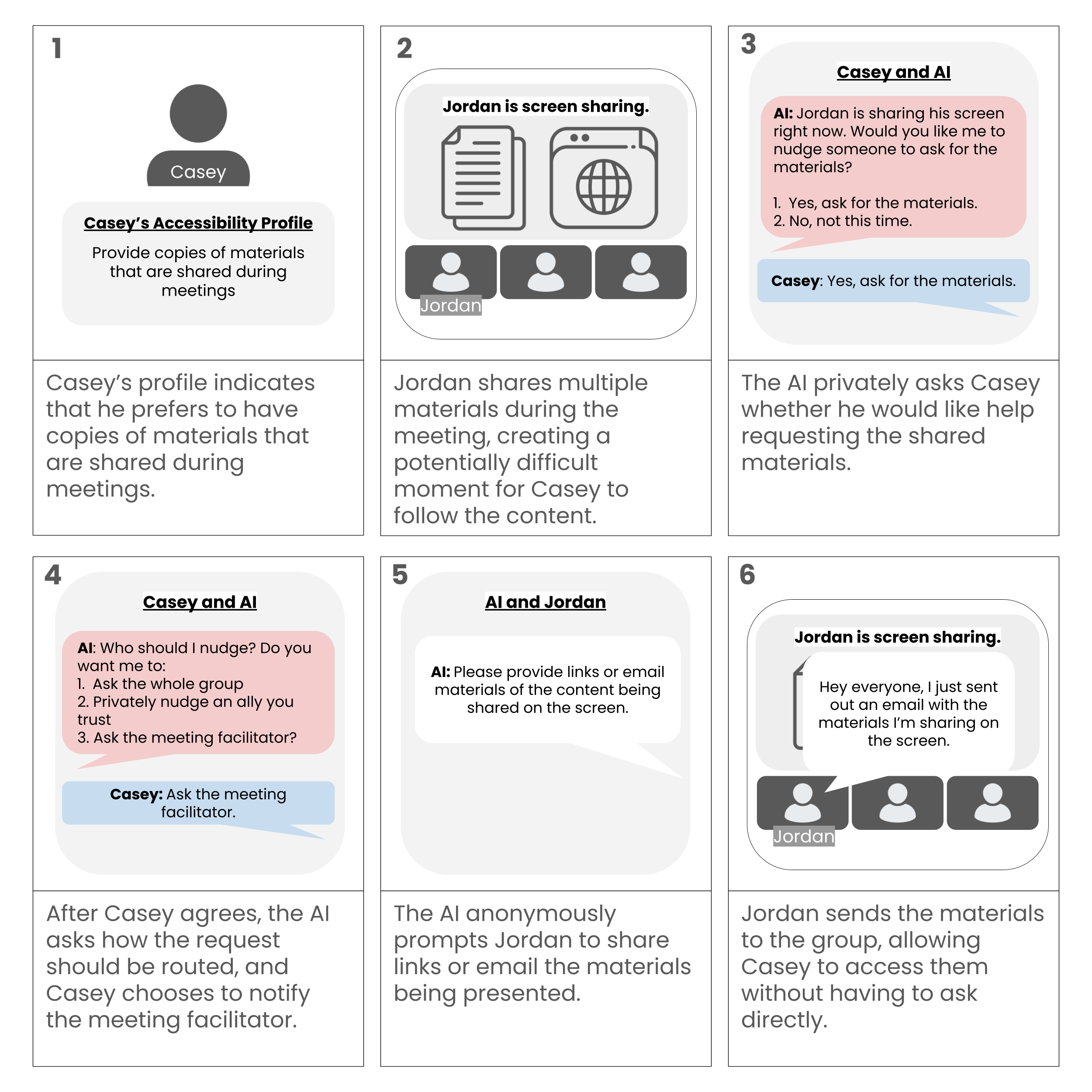}
    \caption{Scenario 2: Anonymous Support Request.}
    \Description{Six-panel storyboard with numbered panels, screen-sharing visuals, private AI prompts, chat-style message bubbles, and captions beneath each panel. Casey's profile indicates a preference for receiving copies of shared materials. While Jordan is screen sharing, the AI privately asks Casey whether he wants help requesting the materials. Casey agrees and chooses to route the request through the facilitator. The AI sends the request, and Jordan shares the materials with the group. The panel captions describe the step-by-step process of AI-mediated, lower-risk support requesting.}\label{fig:appendix_scenario2}
\end{figure*}
\begin{figure*}[htbp]
    \centering
    \includegraphics[width=0.85\textwidth]{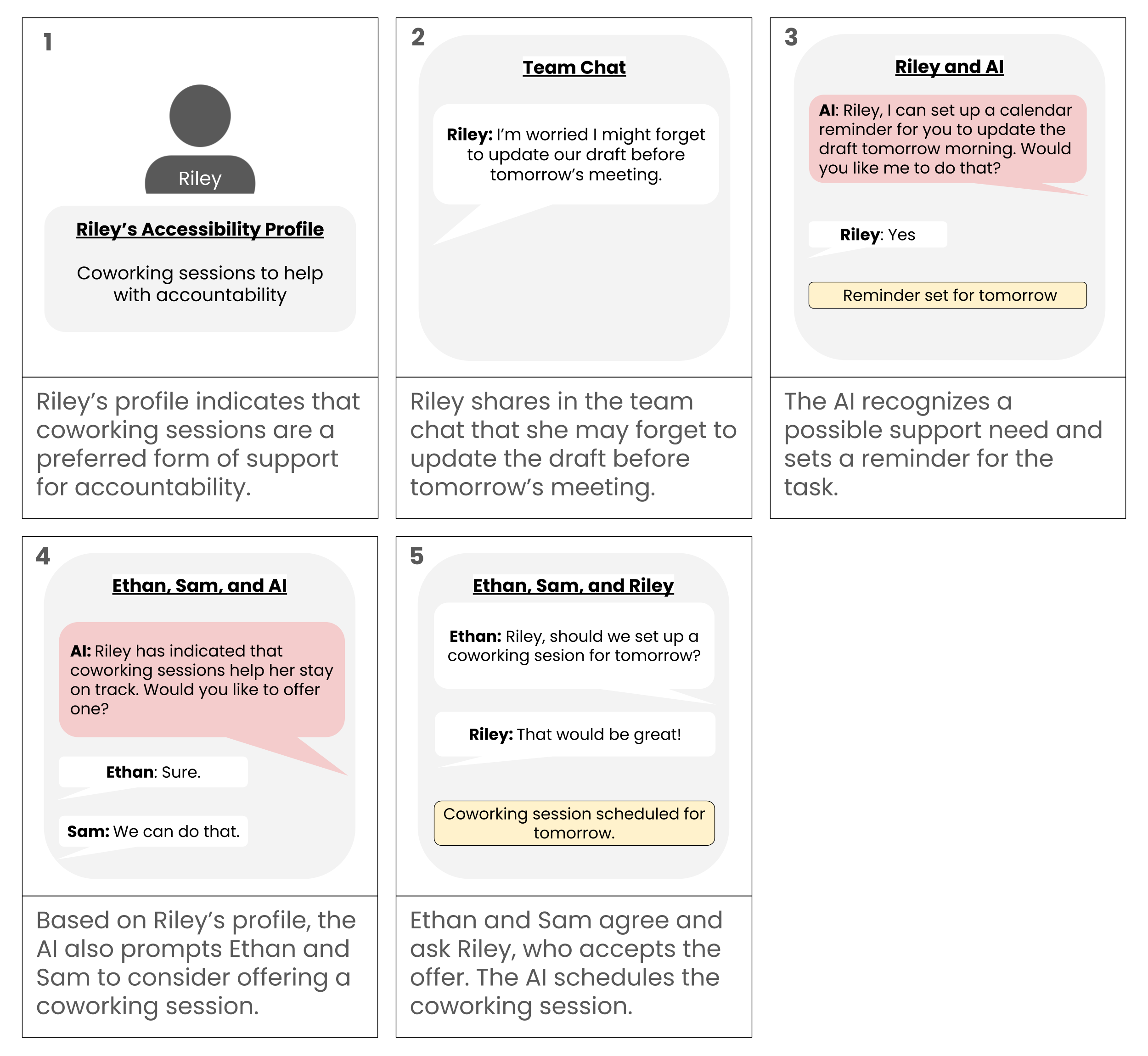}
    \caption{Scenario 3: Support Opportunity Prompts.}
    \Description{Five-panel storyboard with numbered panels, accessibility-profile visuals, team chat and AI message bubbles, highlighted reminder and scheduling boxes, and captions beneath each panel. Riley's profile indicates that coworking sessions help with accountability. Riley says she may forget to update a draft before tomorrow's meeting. The AI offers and sets a reminder, then prompts Ethan and Sam to consider offering a coworking session. Riley accepts, and the session is scheduled. The panel captions describe the progression from stored preference, to support need, to AI prompt, to coordinated team support.}\label{fig:appendix_scenario3}
\end{figure*}
\begin{figure*}[htbp]
    \centering
    \includegraphics[width=0.6\textwidth]{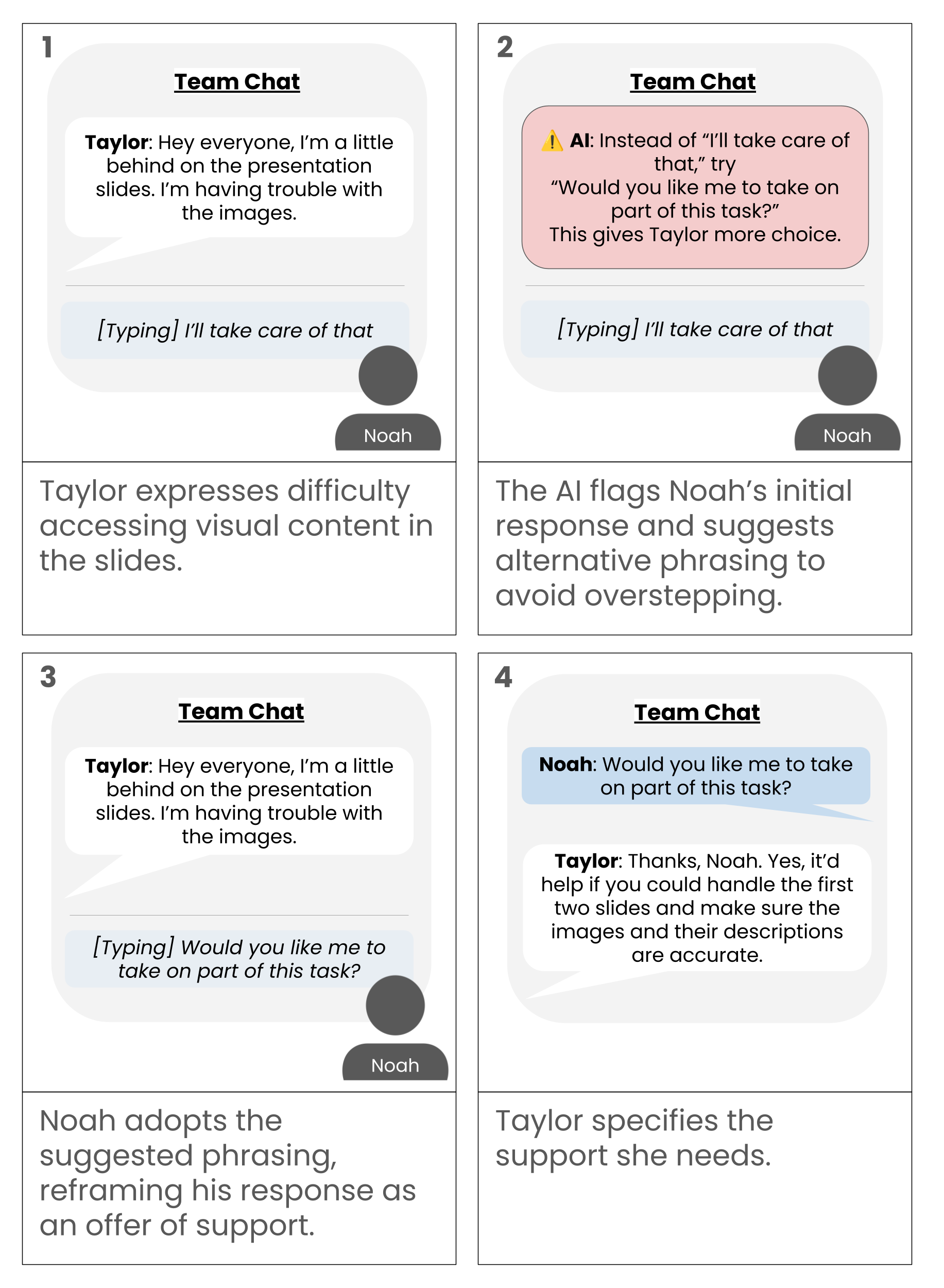}
    \caption{Scenario 4: Phrasing Alternatives for Offering Help.}
    \Description{Four-panel storyboard with numbered panels, chat-style interface, highlighted AI suggestion box, draft and revised message states, and captions beneath each panel. Taylor says she is behind on presentation slides and struggling with the images. The AI flags Noah's initial draft response and suggests a less presumptive alternative. Noah adopts the suggested phrasing, and Taylor responds by specifying the support she needs. The panel captions emphasize the sequence from disclosure, to AI-supported rephrasing, to Taylor's clarification of desired help.}\label{fig:appendix_scenario4}
\end{figure*}

\clearpage
\begin{table*}[!t]
\section{Cross-Case Comparison Matrix}
\label{app:cross_case_matrix}
\centering
\footnotesize
\setlength{\tabcolsep}{4pt}
\renewcommand{\arraystretch}{1.05}
\begin{tabularx}{\textwidth}{p{3.2cm} p{3.4cm} p{4.0cm} X}
\toprule
\textbf{First-Level Theme} & \textbf{Second-Level Theme} & \textbf{Interpretive Codes} & \textbf{Cross-Case Pattern (Summary)} \\
\midrule

\multirow{10}{=}{\textbf{Reducing the Labor of Expressing, Interpreting, and Coordinating Access Needs}}
& \multirow{6}{=}{Addressing Breakdowns in Interpreting Needs and Responses}
& Uncertainty about when and how to help
& Allies hesitated due to lack of clarity about whether support was needed or appropriate. \\

& & Fear of overstepping or being patronizing
& Participants avoided acting to prevent appearing intrusive or condescending. \\

& & Lack of a clear forum for discussing needs
& There were no socially acceptable mechanisms to ask about accessibility needs. \\

& & Misinterpretation from inference
& In the absence of explicit information, participants guessed needs, sometimes incorrectly. \\

& & Gap between awareness and actionability
& Knowing a need exists did not translate into knowing what to do. \\

& & Needs are not always fully knowable
& Participants, especially disabled participants, described difficulty identifying or articulating their own needs. \\
\cmidrule(l){2-4}

& \multirow{4}{=}{Reducing the Burden of Expressing and Coordinating Access Needs}
& Repeated disclosure as labor
& Participants described repeatedly explaining needs as emotionally and cognitively taxing. \\

& & Preference for persistent accessibility profiles
& Storing needs once reduced repeated explanation across changing collaborators. \\

& & Making needs actionable for others
& Explicit preferences enabled more direct and effective support from teammates. \\

& & Preset options signal legitimacy of accommodations
& Predefined options conveyed that accommodations were normalized and acceptable. \\
\bottomrule
\end{tabularx}
\caption{Summarized cross-case analysis matrix (Part 1): reducing the labor of expressing, interpreting, and coordinating access needs.}
\label{app:cross_case_summary_part1}
\end{table*}

\begin{table*}[t]
\centering
\footnotesize
\setlength{\tabcolsep}{4pt}
\renewcommand{\arraystretch}{1.05}
\begin{tabularx}{\textwidth}{p{3.2cm} p{3.4cm} p{4.0cm} X}
\toprule
\textbf{First-Level Theme} & \textbf{Second-Level Theme} & \textbf{Interpretive Codes} & \textbf{Cross-Case Pattern (Summary)} \\
\midrule

\multirow{16}{=}{\textbf{Preserving Human Judgment, Control, and Team Dynamics}}
& \multirow{4}{=}{Distributing the Labor of Allyship}
& Centralization of accessibility labor
& Accessibility work was often concentrated in a single individual, such as a manager or organizer. \\

& & Need to redistribute coordination work
& Participants valued distributing responsibility across teammates or systems. \\

& & Ally capacity constraints
& Even willing allies lacked time, attention, or resources to consistently provide support. \\

& & Role-dependent allyship responsibility
& Willingness to act varied based on role, such as organizer versus collaborator. \\
\cmidrule(l){2-4}

& \multirow{4}{=}{Accounting for Organizational and Social Constraints}
& Professional risk in disclosure
& Participants hesitated to request support due to fear of evaluation, stigma, or retaliation. \\

& & Organizational culture shapes safety
& Tools were only useful when the environment felt safe and supportive. \\

& & Group size affects accountability
& Larger teams diffused responsibility, reducing the likelihood of action. \\

& & Structural constraints on coordination
& Visibility of needs did not guarantee follow-through without clear responsibility. \\
\cmidrule(l){2-4}

& \multirow{4}{=}{Preserving Control, Authenticity, and Human Judgment}
& AI mediation as threat to authenticity
& AI-mediated support risked feeling less genuine or meaningful. \\

& & Loss of control over self-presentation
& Rephrasing tools could misrepresent tone, intent, or identity. \\

& & AI may weaken interpersonal learning
& Overreliance on AI could reduce opportunities to develop allyship skills. \\

& & Value of human intention in support
& Participants emphasized that support should reflect genuine care and judgment. \\
\cmidrule(l){2-4}

& \multirow{4}{=}{Supporting Coordination Without Overstepping Social Boundaries}
& Effort-based boundary for AI involvement
& AI was acceptable for low-effort tasks but not for high-effort or socially complex ones. \\

& & Preference for human negotiation in high-stakes support
& Requests involving time or commitment required direct interpersonal interaction. \\

& & Need for explicit, situational consent
& Participants wanted control over when and how AI acted on their behalf. \\

& & Maintaining agency for both parties
& Both the requester and the ally should retain control over support interactions. \\
\bottomrule
\end{tabularx}
\caption{Summarized cross-case analysis matrix (Part 2): preserving human judgment, control, and team dynamics.}
\label{app:cross_case_summary_part2}
\end{table*}
\clearpage

\end{document}